\documentclass[10pt, journal, a4paper, twocolumn]{IEEEtran}

\usepackage{amsmath,amssymb}
\usepackage{amsfonts,amssymb}
\usepackage{bm}
\usepackage{epsfig}
\usepackage{graphicx}
\usepackage{epstopdf}
\usepackage{psfrag}
\usepackage{cite}
\usepackage{url}
\usepackage{caption}
\usepackage{subcaption}

\newcommand{\y}{\mathbf{y}}
\newcommand{\p}{\mathbf{p}}

\newcommand{\s}{\mathbf{s}}
\newcommand{\n}{\mathbf{n}}
\newcommand{\h}{\mathbf{h}}
\newcommand{\f}{\mathbf{f}}
\newcommand{\Y}{\mathbf{Y}}
\newcommand{\C}{\mathbf{C}}
\newcommand{\R}{\mathbf{R}}
\newcommand{\N}{\mathbf{N}}
\newcommand{\T}{\mathbf{T}}
\newcommand{\W}{\mathbf{W}}
\newcommand{\V}{\mathbf{V}}
\newcommand{\F}{\mathbf{F}}
\newcommand{\I}{\mathbf{I}}
\newcommand{\Q}{\mathbf{Q}}
\newcommand{\U}{\mathbf{U}}
\newcommand{\E}{\mathbf{E}}
\newcommand{\D}{\mathbf{D}}
\newcommand{\mH}{\mathbf{H}}
\newcommand{\mP}{\mathbf{P}}
\newcommand{\cA}{\mathcal{A}}
\newcommand{\cB}{\mathcal{B}}

\newcommand{\rw}{\rightarrow}

\newcommand{\beq}{\begin{equation}}
\newcommand{\eeq}{\end{equation}}
\newcommand{\bbm}{\begin{bmatrix}}
\newcommand{\ebm}{\end{bmatrix}}

\IEEEoverridecommandlockouts

\title{Channel reciprocity calibration in TDD hybrid beamforming massive MIMO systems}

\author{
    Xiwen ˜JIANG, ˜\IEEEmembership{Student Member, ˜IEEE,} 
and ˜Florian ˜Kaltenberger, ˜\IEEEmembership{Member, ˜IEEE}
       \thanks{This work was supported in part by Huawei Mathematical and Algorithmic Sciences Lab in Paris through the project of ``Modeling, Calibrating and Exploiting Channel Reciprocity for Massive MIMO". This work was also supported in part by the French Government (National Research Agency, ANR) through the ``Investments for the Future'' Program \#ANR-11-LABX-0031-01. }
	\thanks{The authors are with Communication Systems Department, EURECOM, Campus SophiaTech, 06410 Biot, France (e-mail: xiwen.jiang@eurecom.fr; florian.kaltenberger@eurecom.fr).}
}  

\begin{document}
\maketitle

\begin{abstract}
%Hybrid analog and digital beamforming structure is a very attractive solution to build low cost massive multiple-input multiple-output (MIMO) systems. Typically these systems use a set of fixed beams since it is not possible to get the channel state for each antenna element individually, both in time division duplex (TDD) and frequency division duplex (FDD) systems. In TDD systems especially, it prohibits the use of reciprocity calibration techniques to acquire full channel state information at the transmitter (CSIT). In this paper, we propose a reciprocity calibration scheme for such hybrid systems that allows to acquire full CSIT in TDD systems with hybrid beamforming structure. Different to existing CSIT acquisition methods, our approach does not require any assumption on the channel model and can obtain near perfect CSIT.
Hybrid analog-digital (AD) beamforming structure is a very attractive solution to build low cost massive multiple-input multiple-output (MIMO) systems. Typically these systems use a set of fixed beams for transmission and reception to avoid the need to obtain channel state information at transmitter (CSIT) for each antenna element individually. However, such a method can not fully exploit the potential of hybrid AD beamforming systems. Alternatively, CSIT can be estimated by assuming a model for the propagation channel, whereas this model is only validated in millimeter-wave (mmWave) band thanks to its poor scattering nature. In this paper, we focus on time division duplex (TDD) systems with hybrid beamforming structure and propose a reciprocity calibration scheme that allows to acquire full CSIT. Different to existing CSIT acquisition methods, our approach does not require any assumption on the channel model and can, in theory, estimate the CSIT up to an arbitrary small error.
\end{abstract}


%=============== Introduction ===============
\section{Introduction}
\label{sec:intro}
Massive multiple-input multiple-output (MIMO) is considered as a key enabler of the next generation of wireless communication networks, as it has the potential to dramatically increase the network capacity \cite{marzetta2010noncooperative}. To bring this concept to practice, it is essential to reduce the cost of building up such complex systems. Among the most promising solutions, hybrid analog-digital (AD) beamforming structure has achieved great attention. By introducing phase shifters and reducing the number of expensive components on digital and RF chains, such as digital-to-analog/analog-to-digital converters (DACs/ADCs) as well as signal mixers, hybrid beamforming structure opens up possibilities to build relatively low cost massive MIMO systems.

A common way of enabling hybrid beamforming is to pre-define a set of fixed beams in the downlink (DL) on which pilots are transmitted to a user equipment (UE) who then simply selects the best beam and sends the index back to the base station (BS), which will use it directly for data transmission \cite{kim2013multi, han2015large}. Such systems have also been specified for LTE-Advanced Pro, in the so-called full dimension (FD) MIMO system\cite{ji2017overview}, but are clearly suboptimal compared to the case where full channel state information at the transmitter (CSIT) is available \cite{flordelis2017massive}. Under the assumption of full CSIT, a hybrid massive MIMO system can achieve the same performance of any fully digital beamforming scheme, as long as the number of RF chains is at least twice the number of data schemes \cite{sohrabi2016hybrid2}. However, acquisition of CSIT in a hybrid massive MIMO system is a non-trivial matter, both for frequency division duplex (FDD) and time division duplex (TDD) systems. 

The problem was studied in the millimeter-wave (mmWave) band in \cite{alkhateeb2014channel}, where the channel can be considered to have only a few number of dominant rays because of the poor scattering nature of the channel. While this method works out well for mmWave, it can hardly be generalized to an arbitrary channel, especially when hybrid beamforming massive MIMO systems are used in a sub-6GHz band. %Additionally, when the channel changes, the whole process of beam training needs to be performed again.
%While these current solutions can make hybrid beamforming work under certain propagation assumptions, neither can acquire a perfect CSIT in a general propagation environment. Thus the potential of hybrid beamforming promised in \cite{el2012low} can be hardly fully released.

In this paper, we propose to perform the DL CSIT acquisition based on the channel reciprocity property in a time division duplex (TDD) system. In fact, as long as the DL and uplink (UL) transmission happens within the channel coherence time, the physical channel is reciprocal. This property was used when massive MIMO was introduced in \cite{marzetta2010noncooperative} to avoid large channel feedback to the BS in UL. The only problem is that the transmit and receive radio frequency (RF) chains in transceivers (hardware from DAC to antenna at the transmit path and that from antenna to ADC at the receive path) are not reciprocal, thus calibration is needed to compensate the hardware asymmetry.

\begin{figure*}[t]
\centering
\includegraphics[width=1.6\columnwidth]{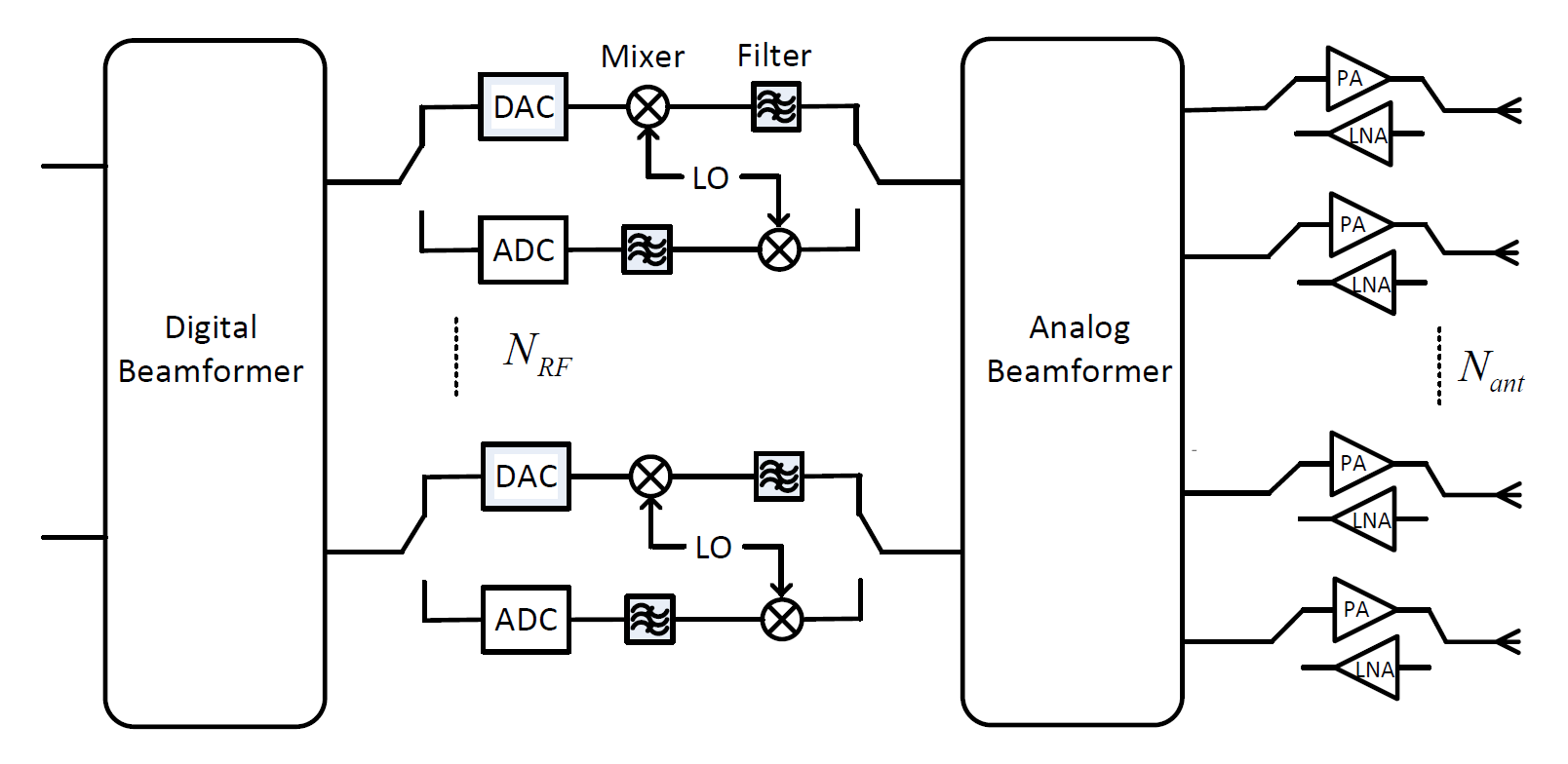}
\caption{The structure of a TDD hybrid beamforming transceiver, where both the transmit and receive paths are shown. The transceiver can dynamically change the connexion of different switches to set itself to the transmitting or receiving mode.}
\label{fig:hybrid_architecture}
\vspace{-1.0em}
\end{figure*}

In a fully digital TDD system, numerous calibration methods have been proposed.  Reciprocity calibration using ``over-the-air" signal processing was introduced for classical MIMO systems in \cite{guillaud2005practical, kaltenberger2010relative}, where the BS exchanges pilots with the UE to estimate bi-directional channel. A total least squares (TLS) problem can then be formulated to estimate the calibration coefficients. Such methods, however, can not be directly applied to massive MIMO systems since all UEs need to feed back their measured DL CSI during the calibration procedure. In \cite{shi2011efficient}, authors propose to choose a reference UE to assist the BS to calibrate in order to reduce the UL feedback.  BS internal calibration was then introduced in \cite{shepard2012argos} for the Argos massive MIMO testbed, where the reference UE is replaced by a reference antenna, so that BS can perform internal calibration without the involvement of UE.  %The problem of Argos reciprocity calibration is that the algorithm is quite sensitive the placement of the reference antenna. 
However, the Argos calibration method is quite sensitive to the placement of the reference antenna, thus is not easy for real environment deployment and not suitable for antennas in a distributed topology.
In order to take up this challenge, methods based on bi-directional transmissions between antenna pairs are proposed in \cite{rogalin2014scalable, vieira2014reciprocity}. These methods were initially designed for distributed massive MIMO systems but show good performance in co-localized systems as well. Other methods such as \cite{papadopoulos2014avalanche, luo2015robust, wei2016mutual} address different aspects in the calibration problem, including speeding up the whole calibration procedure, reducing the number of transmission needed for calibration and calibration dedicated to maximum ratio transmission (MRT). In \cite{vieira2017reciprocity}, the authors propose a maximum likelihood (ML) estimator to enhance the accuracy of reciprocity calibration.

Regardless the variety of calibration methods, none of them can be directly used in a hybrid AD beamforming structure \cite{han2015large}. This is the main reason why TDD reciprocity based methods have been left behind in this type of massive MIMO systems. In this paper, we introduce a reciprocity calibration method which allows us to avoid beam training or selection and acquire CSIT without any assumption on the channel. The main contributions of this paper are as follows:
\begin{itemize}
\item We propose a reciprocity calibration method for TDD hybrid beamforming massive MIMO systems. This problem was never addressed before, although calibration methods for fully digital systems were introduced more than a decade ago.
\item Based on reciprocity calibration, we illustrate that TDD hybrid beamforming systems have the potential to acquire the CSIT up to an arbitrary small error, thus can fully release its beamforming potential. This provides novel ways to operate hybrid beamforming systems rather than performing beam training using a fix set of pre-determined beams.
\end{itemize}

The notation adopted in this paper conforms to the following convention. Vectors and matrices are denoted in lowercase bold and uppercase bold respectively: $\mathbf{a}$ and $\mathbf{A}$. $(\cdot)^*$, $(\cdot)^T$, $(\cdot)^H$, $(\cdot)^{-1}$, $(\cdot)^{-T}$ denote element-wise complex conjugate, transpose, Hermitian transpose, inverse, and transpose together with inverse, respectively. $\otimes$ denotes the Kronecker product operator. $\mbox{diag}\{a_1, a_2, \dots, a_M\}$ denotes a diagonal matrix with its diagonal composed of $a_1, a_2, \dots, a_M$, $\mbox{rank}(\mathbf{A})$ represents the rank of matrix $\mathbf{A}$, whereas $\mbox{vec}(\mathbf{A})$ denotes the vectorization of the matrix $\mathbf{A}$. $\mathbb{C}$ denotes the set of complex numbers. Finally, the Frobenius norm is denoted by $\|\cdot\|_F$.

%------------------------------------------------
\section{System model} 
%------------------------------------------------
\subsection{Hybrid Structure}
\label{subsec:hybrid_structure}
%------------------------------------------------
%\subsection{Hybrid Beamforming Structure} 
%------------------------------------------------
The structure of a TDD hybrid beamforming transceiver is shown in Fig.~\ref{fig:hybrid_architecture} \cite{li2016mmwave} where the digital beamformer is connected to $N_{RF}$ RF chains, which, through an analog beamforming network, are connected with power amplifiers (PAs)/low noise amplifiers (LNAs) and $N_{ant}$ antennas, where $N_{ant} \geq N_{RF}$. Note that it is also possible to place PAs and LNAs in the RF chains before the analog beamformer so that the number of amplifiers are less. However, in that case, each amplifier needs more power since it amplifies signal for multiple antennas. Additionally, in the transmission mode, the insertion loss of analog precoder working in the high power region makes the transceiver less efficient in terms of power consumption. In the reception mode, the fact of having phase shifters before LNAs also results in a higher noise figure in the receiver. It is thus a better choice to have PAs and LNAs close to antennas. To this reason, we stick our study in this paper to the structure in Fig.~\ref{fig:hybrid_architecture}. The discussion in this paper, however, can also be applied to the case where the PAs/LNAs are placed before the analog beamformer.

\begin{figure}[t]
\centering
\begin{subfigure}{.49\columnwidth}
  \centering
  \includegraphics[width=0.99\columnwidth]{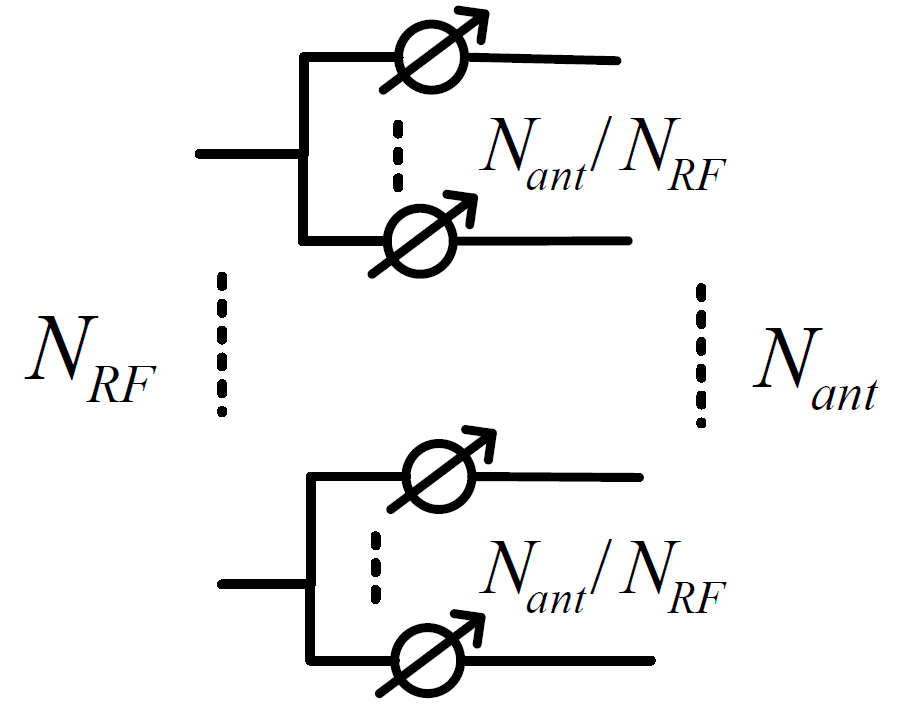}
  \caption{Subarray architecture.}
  \label{fig:analog_per_rf}
\end{subfigure}%
\begin{subfigure}{.49\columnwidth}
  \centering
  \includegraphics[width=\columnwidth]{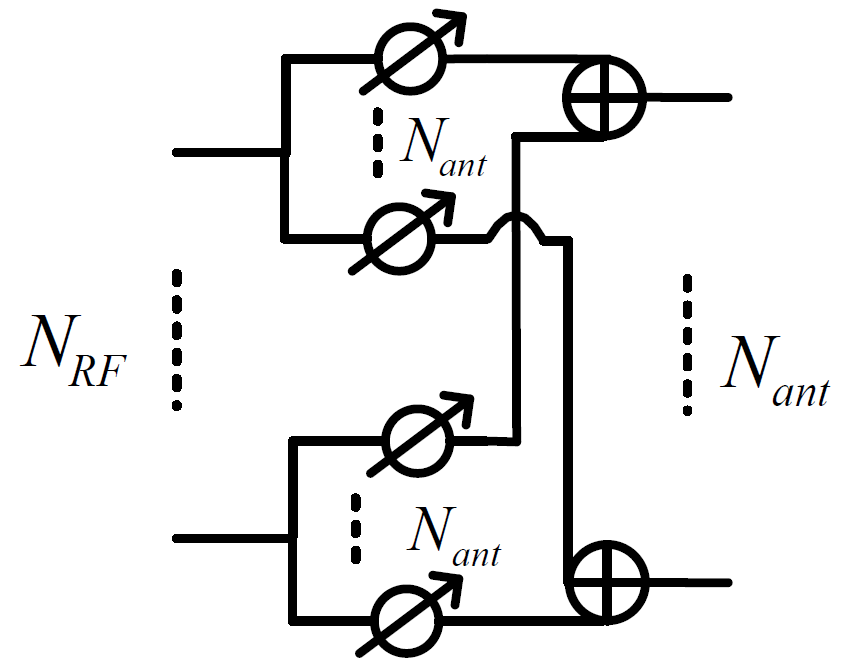}
  \caption{Fully connected.}
  \label{fig:analog_full_str}
\end{subfigure}
\caption{Two types of analog beamforming structure.}
\label{fig:analog_bf_2types}
\vspace{-1.0em}
\end{figure}

The analog beamformer is interpreted as analog precoder and combiner in the transmit and receive path, respectively. Two types of architecture can be found in literature \cite{sohrabi2016hybrid, han2015large}:
\begin{itemize}
 \item \textbf{Subarray architecture:} Each RF chain is connected to $N_{ant}/N_{RF}$ phase shifters as shown in Fig.~\ref{fig:analog_per_rf}. Such a structure can be found in \cite{huang2010hybrid, guo2012hybrid, kim2013multi, roh2014millimeter}
 \item \textbf{Fully connected architecture:} $N_{ant}$ phase shifters are connected to each RF chain. Phase shifters with the same index are then summed up to be connected to the corresponding antenna, as shown in Fig.~\ref{fig:analog_full_str}. This structure can be found in \cite{nsenga2010mixed, el2012low, alkhateeb2013hybrid, alkhateeb2014channel}.
\end{itemize}

In terms of CSIT acquisition, since the BS is not fully digital, assigning orthogonal pilots to different antennas for DL channel estimation per antenna can not be used. Additionally, even assuming that we can have perfect channel estimation for all antennas at the UE, it is unfeasible to feed this information back to the BS, because in a massive MIMO system, the UL overhead will be so heavy that at the time the BS gets the whole CSIT, the information has already outdated. 

In order to address the problem, we make use of the inherent reciprocity property in TDD systems. We firstly show how this is possible for ``subarray architecture'' by enabling reciprocity calibration. We then provide some ideas to calibrate a fully connected hybrid architecture.

\begin{figure*}[t]
\centering
\includegraphics[width=1.6\columnwidth]{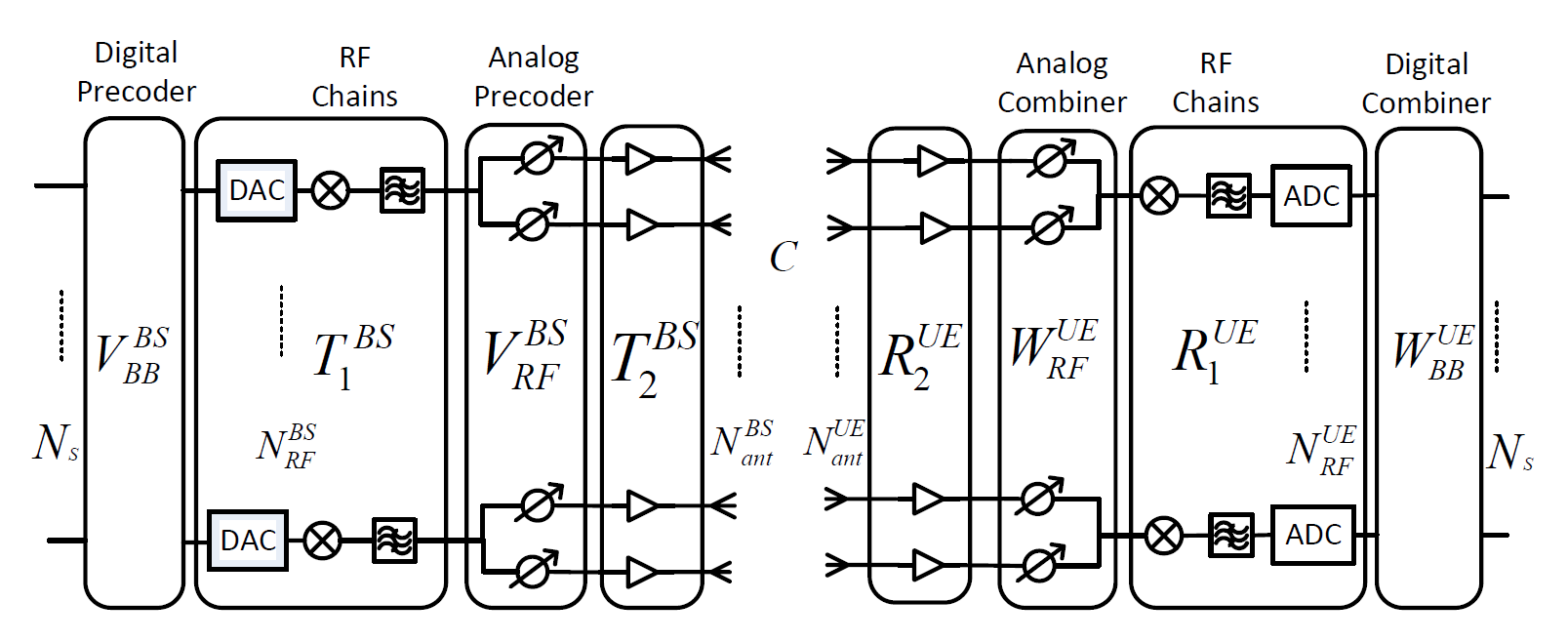}
\caption{Hybrid beamforming system where a BS (left) is transmitting $N_s$ data schemes in the DL to a UE (right). The switches in the BS are connected to the transmit path whereas those in the UE are connected to the receive path.}
\label{fig:hybrid_str}
\vspace{-1.34em}
\end{figure*}

%------------------------------------------------
\subsection{System Model} 
%------------------------------------------------
Consider a subarray hybrid beamforming system with a single user shown in Fig.~\ref{fig:hybrid_str}, where a BS with $N_{ant}^{BS}$ antennas communicates $N_s$ data streams in the DL to a UE with $N_{ant}^{UE}$ antennas. $N_{RF}^{BS}$ and $N_{RF}^{UE}$  are the number of RF chains at the BS and the UE, respectively, such that $N_s \leq N_{RF}^{BS} \leq N_{ant}^{BS}$ and $N_s \leq N_{RF}^{UE} \leq N_{ant}^{UE}$. In Fig.~\ref{fig:hybrid_str}, we use $\V_{BB}^{BS} \in \mathbb{C}^{N_{RF}^{BS}\times N_s}$ and $\W_{BB}^{UE} \in \mathbb{C}^{N_s \times N_{RF}^{UE}}$ to represent the baseband digital beamforming matrix at the BS and at the UE, respectively. $\V_{RF}^{BS} \in \mathbb{C}^{N_{ant}^{BS} \times N_{RF}^{BS}}$ and $\W_{RF}^{UE} \in \mathbb{C}^{N_{RF}^{UE} \times N_{ant}^{UE}}$ are the analog beamforming precoders and combiners. We use $\T_1^{BS} \in \mathbb{C}^{N_{RF}^{BS}\times N_{RF}^{BS}}$, $\T_2^{BS} \in \mathbb{C}^{N_{ant}^{BS}\times N_{ant}^{BS}}$, $\R_1^{UE} \in \mathbb{C}^{{N_{RF}^{UE}\times N_{RF}^{UE}}}$ and $\R_2^{UE} \in \mathbb{C}^{N_{ant}^{UE}\times N_{ant}^{UE}}$ to represent the transfer functions of the corresponding hardwares. The diagonal elements of $\T_1^{BS}$ and $\R_1^{UE}$ capture the hardware characteristics of the $N_{RF}^{BS}$ and $N_{RF}^{UE}$ RF chains including the DACs/ADCs, signal mixers and some other components around, whereas, their off-diagonal elements represent the RF crosstalk. Similarly, the diagonal of $\T_2^{BS}$ and $\R_2^{UE}$ are used to represent the properties of amplifiers as well as some surrounding components after phase shifter on each branch and their off-diagonal elements represent RF crosstalk and antenna mutual coupling \cite{balanis2016antenna}. If we transmit a signal $\s$ through a channel $\C \in \mathbb{C}^{N_{ant}^{UE}\times N_{ant}^{BS}}$, at the output of the UE's digital combiner, we have
\begin{equation}
\label{eqn:sig_model}
\y = \W_{BB}^{UE}\R_1^{UE}\W_{RF}^{UE}\R_2^{UE}\C\T_2^{BS}\V_{RF}^{BS}\T_1^{BS}\V_{BB}^{BS}\s + \n,
\end{equation}
where $\y$ is the $N_s\times 1$ received signal vector and $\n$ following the circularly symmetric complex Gaussian distribution $\mathcal{CN}(0,\sigma_n^2\I)$ is the noise vector. 

In a TDD system, the physical channel is reciprocal within the channel coherence time, i.e., in the reverse transmission, the UL physical channel from UE to BS can be represented by $\C^T$. However, apart from special cases in mmWave, where the channel enjoys limited scattering property, channel estimation, usually performed in the digital domain, does not directly provide $\C$ but a composite channel including $\C$ and transceivers' hardware properties, which are not reciprocal. This will be detailed in Section~\ref{subsec:csit_aquis}.

%------------------------------------------------
\section{CSIT acquisition based on TDD reciprocity calibration} 
%------------------------------------------------
In this section, we describe how to acquire accurate CSIT based on TDD channel reciprocity. Especially, as transmit and receive RF chains break down the inherent reciprocity, we introduce our calibration method to compensate the hardware asymmetry. 

%------------------------------------------------
\subsection{Equivalent System Model} 
%------------------------------------------------
In order to acquire CSIT and calibrate TDD systems, let us firstly introduce an equivalent system model which simplifies the signal model in \eqref{eqn:sig_model}, where we observe that the hardware blocks are mixed up with digital and analog beamforming matrices. Note that $\T_1^{BS}$ (similar for $\R_1^{UE}$) represents the hardware properties on the $N_{RF}^{BS}$ RF chains, where the $N_{RF}^{BS}$ diagonal elements mainly capture the random phases generated by the corresponding RF chains and the off-diagonal elements represent the RF crosstalk, i.e., the RF leakage from one RF chain to the others. Proper RF circuit design usually ensures very small RF crosstalk with regard to the diagonal values, which is also proven by the measurement results in \cite{Jiang2015}, leading to the fact that, in reality, $\T_1^{BS}$ and $\R_1^{UE}$ can be considered to be diagonal. Since $\V_{RF}^{BS}$ and $\W_{RF}^{UE}$, representing the analog beamformers for each RF chain, have block diagonal structures, the matrix multiplication is commutative if we introduce a Kronecker product such as $\V_{RF}^{BS}\T_1^{BS} = (\T_1^{BS}\otimes \I_{BS})\V_{RF}^{BS}$ and $\R_1^{UE}\W_{RF}^{UE} = \W_{RF}^{UE}(\R_1^{UE}\otimes \I_{UE})$, where $\I_{BS}$ and $\I_{UE}$ are identity matrices of size $N_{ant}^{BS}/N_{RF}^{BS}$ and $N_{ant}^{UE}/N_{RF}^{UE}$, respectively. The signal model in \eqref{eqn:sig_model} thus has an equivalent representation as
\begin{equation}
\begin{aligned}
\y = & \underbrace{\W_{BB}^{UE}\W_{RF}^{UE}}_{{\W}_{UE}}
\underbrace{(\R_1^{UE}\otimes \I_{UE})\R_2^{UE}}_{{\R}_{UE}} \C\\
& \underbrace{\T_2^{BS}(\T_1^{BS}\otimes \I_{BS})}_{{\T}_{BS}}
\underbrace{\V_{RF}^{BS}\V_{BB}^{BS}}_{{\V}_{BS}}\s + \n,
\label{eqn:sig_equ_model}
\end{aligned}
\end{equation}
where we group up the digital and analog transmit and receive beamforming matrices into $\V_{BS}$ and $\W_{UE}$. The hardware transfer functions are merged to $\T_{BS}$ and $\R_{UE}$.

\begin{figure}[t]
\centering
\includegraphics[width=0.7\columnwidth]{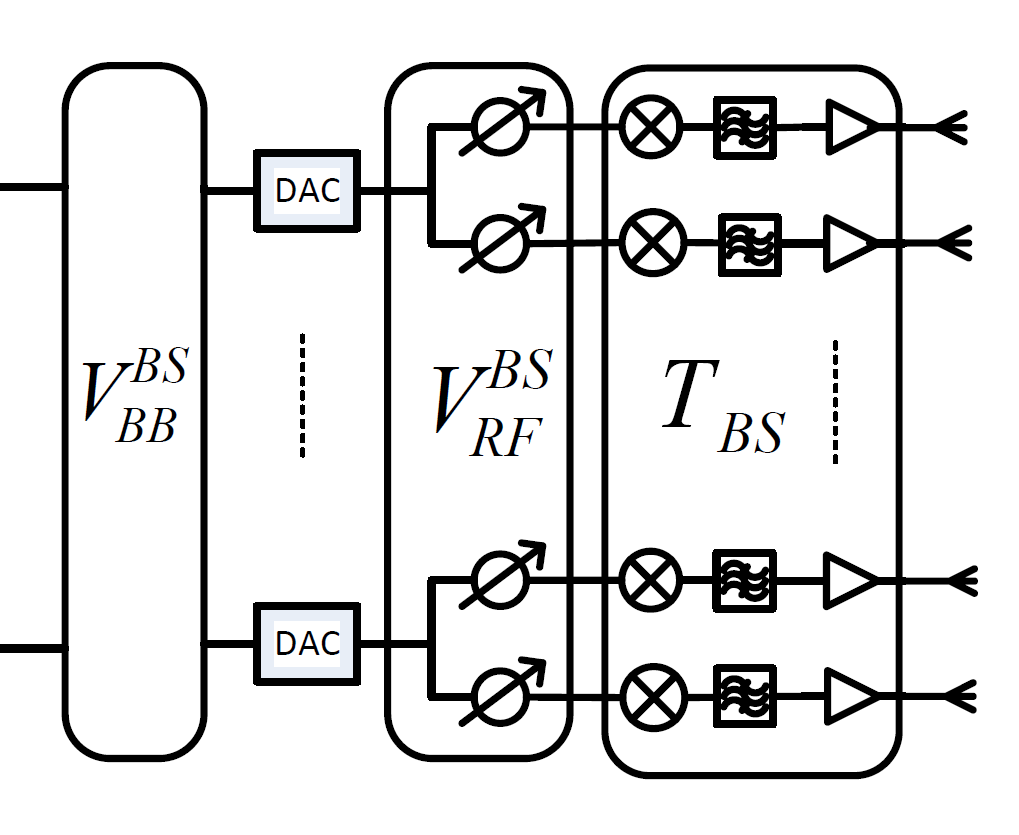}
%\caption{Equivalent hybrid structure where shared hardware components (mixers, filters) on RF chain are copied on each branch with phase shifters. The hardware components are then re-ordered such that all components in $\T_{BS}$ go to the front end near the antennas.}
\caption{Equivalent hybrid structure.}
\label{fig:hybrid_str3}
\vspace{-1.34em}
\end{figure}

An intuitive understanding of this alternative representation on the BS transmit part is shown in Fig.~\ref{fig:hybrid_str3}, where we
1) replace all shared hardware components (mixers, filters) on RF chain by its replicas on each branch with a phase shifter;
2) change the order of hardware components such that all components in $\T_{BS}$ go to the front end near the antennas.

Note that this equivalent model is general for different hardware implementation, i.e., no matter how hardware impairments are distributed on the hybrid structure, we can always use these two steps to create an equivalent system model. For example, if there's any hardware impairment within the phase shifter or in DAC, they can also be extracted out and put into $\T_{BS}$ using the same methodology.

%------------------------------------------------
\subsection{Full CSIT Acquisition Based on Reciprocity Calibration} 
\label{subsec:csit_aquis}
Let us look at the DL and UL transmission between the BS and the UE using TDD mode, the bi-directional transmission represented in the equivalent signal model is given by:
\begin{equation}
\left\{
\begin{aligned}
\y_{BS\rw UE} &= \W_{UE}\underbrace{\R_{UE}\C\T_{BS}}_{\mH_{BS\rw UE}}\V_{BS}{\s}_{BS} +\n_{UE},\\
\y_{UE\rw BS} &= \W_{BS}\underbrace{\R_{BS}\C^T\T_{UE}}_{\mH_{UE\rw BS}}\V_{UE}{\s}_{UE} + \n_{BS},\\
\end{aligned}
\right.
\end{equation}
where $\C$ and $\C^T$ are the reciprocal DL and UL air propagation channel. From the point of view of digital signal processing, the channel does not only include the physical channel $\C$ but also the hardware transfer functions of the radio front ends, thus we define the effective channel $\mH_{BS\rw UE} = \R_{UE}\C\T_{BS}$ and $\mH_{UE\rw BS}=\R_{BS}\C^T\T_{UE}$. Since the transmit and receive RF chains use different hardware components, it is clear that $\T_{BS} \neq \R_{BS}^T$ and $\T_{UE} \neq \R_{UE}^T$. Thus the hardware radio front ends break the TDD channel reciprocity, i.e. $\mH_{BS\rw UE}\neq \mH_{UE\rw BS}$.

In order to compensate the hardware asymmetry and to achieve the reciprocity, we establish the relationship between the DL and UL effective channels as follows
\begin{equation}
\mH_{BS\rw UE} =\R_{UE}\T_{UE}^{-T}\mH_{UE\rw BS}^T\R_{BS}^{-T}\T_{BS}.
\end{equation}
Defining $\F = \R^{-T}\T$ for both BS and UE, we have 
\begin{equation}
\mH_{BS\rw UE} =\F_{UE}^{-T}\mH_{UE\rw BS}^T\F_{BS}.
\label{eqn:H_recip}
\end{equation}
We observe that the DL CSIT can be represented as the UL CSI $\mH_{UE\rw BS}$ tuned with two matrices only dependent on the transceivers' hardware, denoted as $\F_{BS}$ and $\F_{UE}$, which are named as calibration matrices at the BS and the UE, respectively. As long as we have the three matrices in \eqref{eqn:H_recip}, we can estimate the DL CSIT.

Note that if the UE has only one antenna, $\F_{UE}$ becomes a scalar and can be ignored, since the ambiguity of a complex scalar value on the obtained CSIT will not change the final created beam pattern \cite{shepard2012argos}. 
%Otherwise, the UE can report back its calibration matrix or the BS can simply consider $\F_{UE}$ as an identity matrix without loosing too much performance in most cases.
According to \cite{R1-091752, R1-091794, R1-094622}, even if the UE has more than one antenna (but significantly less than the eNB), the UE calibration error has little effect to the performance of reciprocity. Especially phase calibration errors at the UE have no effect on the performance, and relative amplitude calibration mismatch at UE side can have some impact. Thus, when the antenna number at the UE is limited, the UE side calibration is not necessarily needed from the point of view of DL beamforming. Taking $\F_{UE}$ as the identity matrix does not impact much the performance\footnote{In a multi-user scenario, the impact from UE side calibration might increase with the number of served UEs. In this case, each UE can feed back its calibration coefficients back to the BS.}.

It is worth noting that both $\F_{BS}$ and $\F_{UE}$ represent hardware properties, which are independent to the propagation channel $\C$, leading to the fact that they are quite stable during the time. Measurements in \cite{shepard2012argos} show that the variation of calibration coefficients deviates from the mean angle with an average of less than 2.6\% (maximum 6.7\%), and from the mean amplitude less than 0.7\% (maximum 1.4\%), over a period of 4 hours. This implies that calibration does not have to be performed very frequently. %Additionally, feeding back the UE calibration coefficients to the BS is feasible in terms of UL resource consumption regarding to this long period.

In the sequel, we firstly describe the effective channel estimation method for $\mH_{UE\rw BS}$ estimation. We then present an internal reciprocity calibration scheme, where BS (or UE if needed) can estimate its own calibration matrices internally.

%------------------------------------------------
\subsection{Effective Channel Estimation} 
%------------------------------------------------
\label{subsec:eff_ch_est}
In order to obtain the UL CSI, we need to estimate the effective channel based on pilot transmission. This is also needed for internal calibration at the BS and UE side, thus in order to make the description general, we drop the subscript BS and UE and use $\mH=\R\C\T$ to denote the effective channel, where $\T$ and $\R$ are $N_{ant}^t \times N_{RF}^t$ and $N_{RF}^r \times N_{ant}^r$ matrices.
 
Consider sending pilots ($\s=\p$) using $K$ transmit precoders combined with $L$ different receive combiners, we can totally accumulate $KL$ measurements:
\begin{equation}
\underbrace{[\y_{l,k}]}_{\Y} = 
\underbrace{[\W_1^T, \dots, \W_L^T]^T}_{\tilde{\W}}\mH\underbrace{[\V_1\p_1, \dots, \V_K\p_K]}_{\tilde{\mP}} 
+ \underbrace{[\n_{l,k}]}_{\N}.
\end{equation}
where $\y_{l,k}$ is the block element of $\Y$ on the $l^{th}$ row and $k^{th}$ column. $\tilde{\W}$ and $\tilde{\mP}$ are matrices of size $N_sL\times N_{ant}^r$ and $N_{ant}^t \times K$, respectively. To obtain the channel estimation, we vectorize the receive vector as 
\begin{equation}
\mbox{vec}(\Y) =\underbrace{\tilde{\mP}^T\otimes\tilde{\W}}_{\D}\cdot\mbox{vec}(\mH) + \mbox{vec}(\N),
\end{equation}
where we define $\D=\tilde{\mP}^T\otimes\tilde{\W}$. The least squares (LS) channel estimator is
\begin{equation}
\mbox{vec}(\mH) = (\D^H\D)^{-1}\D^H\cdot\mbox{vec}(\Y).
\label{eqn:ch_est}
\end{equation}
In order to guarantee that the estimation problem is over determined, we should have $\mbox{rank}(\D) \geq N_{ant}^t\times N_{ant}^r$, where $\mbox{rank}(\D) = \mbox{rank}(\tilde{\mP}^T)\mbox{rank}(\tilde{\W})$ according to Kronecker product's property on matrix rank. Noting that $\mbox{rank}(\tilde{\mP}^T) \leq \min (N_{ant}^t, K)$ and $\mbox{rank}(\tilde{\W}) \leq \min (N_sL, N_{ant}^r)$, 
thus, in order to meet the sufficient condition of over determination on the estimation problem, we should have $K \geq N_{ant}^t$ and $L \geq N_{ant}^r/N_s$.
%also noting that $\mbox{rank}(\tilde{\mP}^T) \leq  N_{ant}^t$ and  $\mbox{rank}(\tilde{\W})) \leq  N_{ant}^r$, we can conclude that  $\mbox{rank}(\tilde{\mP}^T) =  N_{ant}^t$ and $\mbox{rank}(\tilde{\W}) = N_{ant}^r$. Thus, $K \geq N_{ant}^t$ and $L \geq N_{ant}^r/N_s$.

Note that since the objective here is to estimate the effective channel, digital precoder and combiner are not necessarily needed, i.e. pilots for channel estimation can be inserted after the digital precoder. In this case  $N_s=N_{RF}$ and $L \geq N_{ant}^r/N_{RF}$. Additionally, in a multi-carrier system, where, for example, orthogonal frequency division multiplexing (OFDM) modulation is used, it is possible to allocate different carriers to the pilots of different RF chains. Assuming $\beta$ the number of frequency multiplexing factor on transmit RF chains, the number of the needed transmit precoder $K \geq N_{ant}^t/\beta$.

%In a practical system where orthogonal frequency division multiplexing (OFDM) modulation is used, channel estimation on different RF chains can be performed on orthogonal time-frequency resource elements, similar as the current LTE system. Additionally, since the objective is to estimate the effective channel, digital precoder and combiner are not necessarily needed, i.e. pilots for channel estimation can inserted after the digital precoder. Thus, on a given time-frequency resource element allocated for channel estimation, we usually have $N_s = N_{RF}^t = 1$, thus the needed number of analog precoder and combiner would be merely $K \geq 1$ and $L\geq N_{ant}^r/N_{RF}^r$.
%For example, if $N_{RF}^r = N_{RF}^t = N_{RF}$, we can take $N_s=N_{RF}$ and $V_{BB}=W_{BB}=\I_{RF}$, so that the pilots are only precoded with analog beamformer. 
%In this case, $K\geq N_{ant}^t/N_{RF}$ and $L \geq N_{ant}^r/N_{RF}$.

The effective channel estimation can be used to obtain UL channel estimation but will also be served to estimate calibration matrices as will be presented hereafter.

%------------------------------------------------
\subsection{Internal Reciprocity Calibration} 
\label{subsec:intern_calib}
%------------------------------------------------
One basic idea in estimating calibration matrix consists in accumulating extensively pairs of channel measurement $\hat{\mH}_{BS\rw UE}$ and $\hat{\mH}_{UE\rw BS}$, based on which $\F_{BS}$ and $\F_{UE}$ can be estimated. However, such a method implies that we have to exchange the estimated channel information between BS and UE during the calibration, which introduces an extra cost. It is thus more reasonable to perform calibration internally within the antenna array, at the BS and the UE, independently.

\begin{figure}[!t]
\centering
\includegraphics[width=\columnwidth]{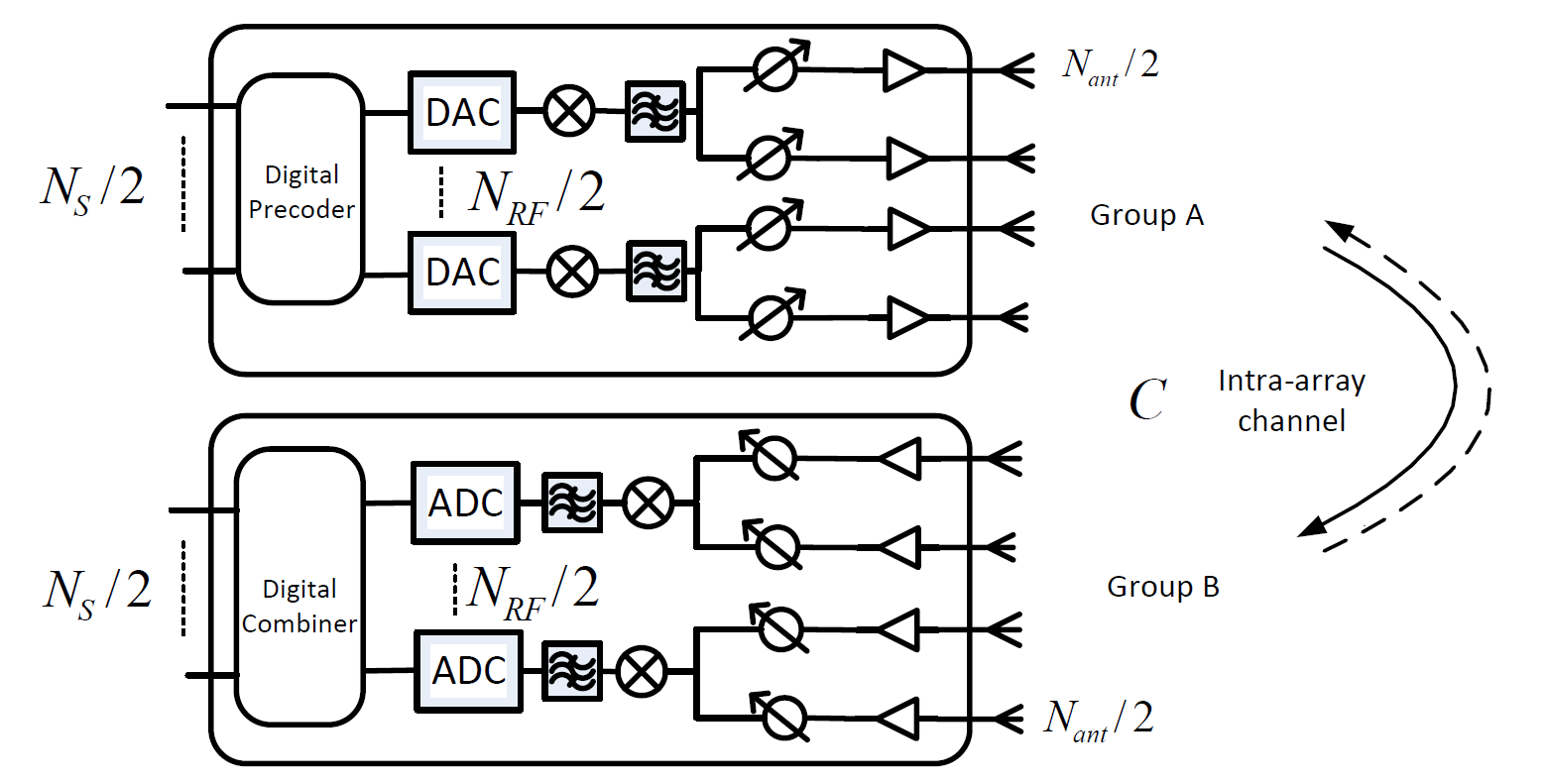}
\caption{Internal calibration where the whole antenna array is partitioned into group $\cA$ and group $\cB$. We then perform intra-array measurement between the two groups.}
\label{fig:inter_calib}
\vspace{-1.0em}
\end{figure}

Internal calibration means that the pilot-based channel estimation happens between different antennas of the same transceiver. Let us equally partition the total $N_{ant}$ antennas into two groups $\cA$ and $\cB$, e.g., $\cA = \{1,2,...,\frac{N_{ant}}{2}\}$ and $\cB = \{\frac{N_{ant}}{2}+1,...,N_{ant}\}$, as shown in Fig.~\ref{fig:inter_calib}. When the antennas in group $\cA$ are connected to the transmit path of $\frac{N_{RF}}{2}$ RF chains, the antennas in group B are connected to the receive path of the rest $\frac{N_{RF}}{2}$ RF chains. We firstly perform an intra-array transmission from $\cA$ to $\cB$, and within the channel coherence time, we switch the roles of group $\cA$ and $\cB$ in order to transmit signal from $\cB$ to $\cA$. The bi-directional received signals are given by
\begin{equation}
\left\{
\begin{aligned}
\y_{\cA\rw \cB} &= \W_\cB\R_\cB\C\T_\cA\V_\cA\p_\cA +\n_{\cA\rw \cB},\\
\y_{\cB\rw \cA} &= \W_\cA\R_\cA\C^T\T_\cB\V_\cB\p_\cB + \n_{\cB\rw \cA},
\end{aligned}
\right.
\label{eqn:bi_trans}
\end{equation}
where $\p_\cA$ and $\p_\cB$ are transmitted pilots, $\C$ is the reciprocal intra-array channel whereas $\n_{\cA\rw \cB}$ and $\n_{\cB\rw \cA}$ are noise. 

If we use $\mH_{\cA\rw \cB}=\R_\cB\C\T_\cA$ and $\mH_{\cB\rw \cA}=\R_\cA\C^T\T_\cB$ to represent the bi-directional effective channels between group $\cA$ and $\cB$, including the physical channel in the air as well as transceiver's hardware, similar to \eqref{eqn:H_recip}, we have
\begin{equation}
\label{eqn:recip}
\mH_{\cA\rw \cB} ={\F_\cB}^{-T}\mH_{\cB\rw \cA}^T{\F_\cA},
\end{equation}
where $\F_\cA = \R_\cA^{-T}\T_\cA$ and $\F_\cB = \R_\cB^{-T}\T_\cB$ are the calibration matrices. In a practical system, the off-diagonal elements $\F_\cA$ and $\F_\cB$ representing the RF crosstalk and antenna mutual coupling are much smaller than the diagonal elements representing the main calibration coefficients. In fact, in-depth theoretical modeling on the calibration matrix in \cite{petermann2013multi}, system measurements from experiment such as in \cite{Jiang2015}, as well as practical experience in fully digital testbeds such as in \cite{shepard2012argos, vieira2017reciprocity} all indicate tha $\F_\cA$ and $\F_\cB$ can be considered to be diagonal. The calibration estimation problem will thus be much simplified. Besides if we use $\F$ to denote the whole calibration matrix, we have $\F = \mbox{diag}\{\F_\cA, \F_\cB\} = \mbox{diag}\{f_1,\ldots,f_N\}$, where $\mbox{diag}\{\cdot\}$ represents the operation to construct a diagonal matrix with given elements on its diagonal. 

Internal reciprocity calibration consists in estimating $\F$ based on the intra-array channel measurement $\hat{\mH}_{\cA\rw\cB}$ and $\hat{\mH}_{\cB\rw \cA}$, without any involvement of other transceivers. Since the calibration coefficients stay quite stable during a relatively long time, once they are estimated, we can use them together with instantaneously estimated UL channel estimation $\hat{\mH}_{UL}$ to obtain CSIT. 

Let us denote the antenna index in group $\cA$ and $\cB$ by $i$ and $j$, respectively, since $\F$ is a diagonal matrix, we have
\begin{equation}
\begin{split}
&h_{i\rw j} =  f_{j}^{-1}h_{j\rw i}f_{i}, \\
\mbox{where,} \;\; i\in \{1,2,& ...,\frac{N_{ant}}{2}\}, j\in \{\frac{N_{ant}}{2}+1,...,N_{ant}\}.
\end{split}
\end{equation}
%The LS estimation problem is thus given by,
%\begin{equation}
%\{f_i, f_j \}=  \mbox{argmin} \sum_{i\in A, j\in B}|f_j h_{i\rw j} -f_i h_{j\rw i}|^2.
%\end{equation}
%The optimization problem can be solved by setting the partial derivatives of the cost function with regard to $f_i^*$ and $f_j^*$ to zeros, under a constraint, e.g. $f_1 =1$, same as in \cite{rogalin2014scalable}. The solution is given by
The problem then becomes very similar to that in \cite{rogalin2014scalable}. Let us use $J$ to denote the cost function of a LS problem:
\begin{equation}
J(f_1, f_2,...,f_{ant}) =  \sum_{i\in A, j\in B}|f_j h_{i\rw j} -f_i h_{j\rw i}|^2.
\label{eqn:LS_cost_fun}
\end{equation}

Estimating the calibration coefficients consists in minimizing $J$ subject to a constraint, e.g., assuming a unit norm or the first calibration coefficient to be known. We adopt here the unit norm constraint, such as $\|\f\| = 1$, where $\f$ is the diagonal vector of $\F$. The Lagrangian function of the constrained LS problem is given by
\begin{equation}
L(\f, \lambda) = J(\f) - \lambda(\|\f\|^2-1),
\end{equation}
where $\lambda$ is the Lagrangian multiplier. By setting the partial derivatives of $L(\f, \lambda)$ with regard to $f_i^*$ and $f_j^*$ to zeros, respectively, where $f_i^*$ and $f_i$ are treated as if they were independent variable \cite{hjorungnes2007complex}, we obtain
\begin{equation}
\left\{
\begin{split}
\frac{\partial L(\f, \lambda)}{\partial f_i^*} &= \Sigma_{j\in B} (f_i|h_{j\rw i}|^2 - f_j h_{j\rw i}^*h_{i\rw j}) -\lambda f_i= 0,\\
\frac{\partial L(\f, \lambda)}{\partial f_j^*} &= \Sigma_{i\in A} (f_j|h_{i\rw j}|^2 - f_i h_{i\rw j}^*h_{j\rw i})  -\lambda f_j= 0.
\end{split}
\right.
\label{eqn:partial}
\end{equation}
The matrix representation of \eqref{eqn:partial} is $\Q\f =\lambda\f$, where $\Q\in\mathbb{C}^{{N_{ant}}\times N_{ant}}$, whose element on its $i$-th row and $m$-th column is
\begin{equation}
Q_{i,m} = \left\{ 
\begin{split}
&\Sigma_{j\in \cB}|h_{j \rw i}|^2 \; &\mbox{for} \;\;  &m = i,\\
&-h_{m \rw i}^*h_{i\rw m} \;&\mbox{for}\;\;  &m \in \cB,%\;\; i\in A \;\mbox{and}\; j\in B
%&0 \;&\mbox{for} \;\; &j \neq i, \;\; i\not\in A \;\mbox{or}\; j\not\in B
\end{split}
\right.
\end{equation}
and the element on the $j$-th row and $m$-th column is given by
\begin{equation}
Q_{j,m} = \left\{
\begin{split}
&\Sigma_{i\in \cA}|h_{i \rw j}|^2 \; &\mbox{for} \;\;  &m = j,\\
&-h_{m \rw j}^*h_{j\rw m} \;&\mbox{for}\;\;  &m \in \cA.%\;\; i\in A \;\mbox{and}\; j\in B
%&0 \;&\mbox{for} \;\; &i \neq j, \;\; i\not\in A \;\mbox{or}\; j\not\in B
\end{split}
\right.
\end{equation}
whereas all other elements are $0$. The solution is given by the eigenvector of $Q$ corresponding to the eigenvalue having smallest magnitude.

%Once both BS and UE estimate the calibration coefficients and UE feeds back $\F_{UE}$, the BS can use \eqref{eqn:H_recip} to estimate DL CSIT.

%------------------------------------------------
\subsection{Calibration for fully connected structure} 
\label{subsec:calib_full}
\begin{figure*}[t]
\centering
\includegraphics[width=1.8\columnwidth]{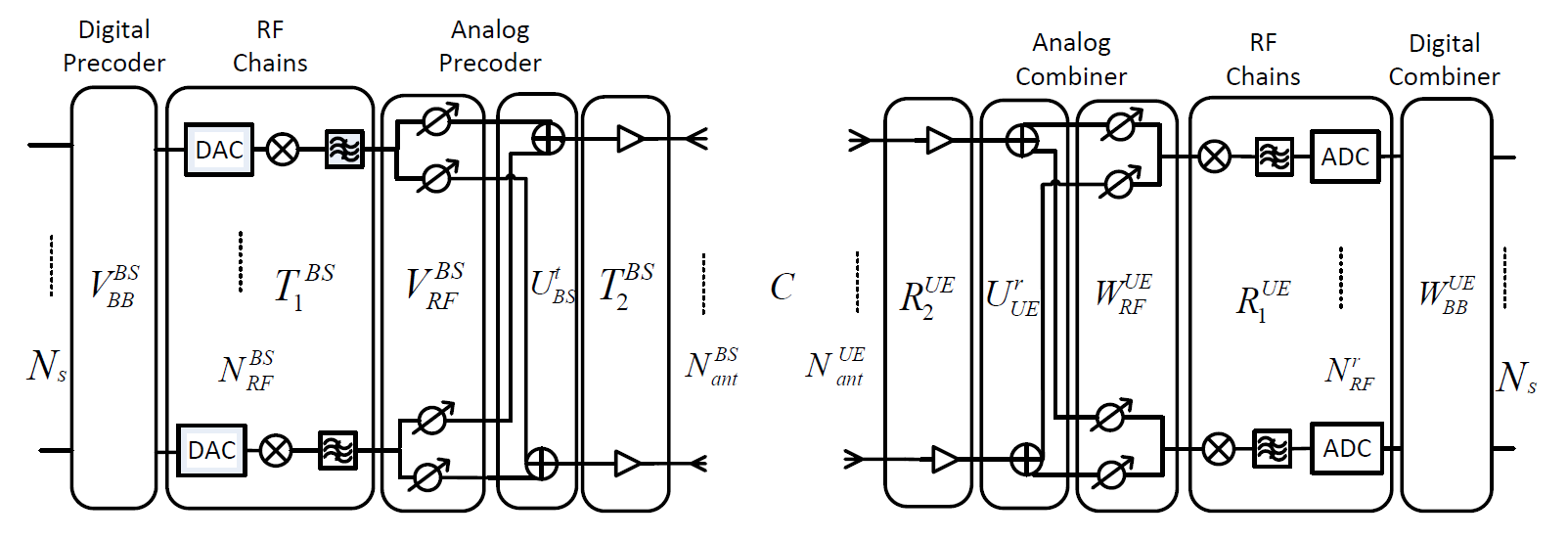}
\caption{Hybrid beamforming system where both the BS and UE have full connected architecture as the analog beamformer}
\label{fig:hybrid_sys_full_str}
\vspace{-1.0em}
\end{figure*}

%-----------------------------------------------
Until now, we have concentrated the reciprocity based CSIT acquisition method under the subarray structure. In this section, we give some ideas on how to calibrate a fully connected architecture for CSIT acquisition. Consider a system with BS and UE both using fully connected hybrid beamforming structure as in Fig.~\ref{fig:hybrid_sys_full_str}.
We use $\U^t_{BS} \in \mathbb{C}^{N_{ant}^{BS}\times N_{ant}^{BS}N_{RF}^{BS}}$  and $\U^r_{UE} \in \mathbb{C}^{N_{RF}^{UE}N_{ant}^{UE}\times N_{ant}^{UE}}$ to denote the summation array between the PA and the antennas at the BS and the corresponding summation operation between the antennas and LNAs at the UE, respectively. The signal model \eqref{eqn:sig_model} can be written as
\begin{equation}
\label{eqn:sig_model_full}
\y = \W_{BB}^{UE}\R_1^{UE}\W_{RF}^{UE}\U^r_{UE}\R_2^{UE}\C\T_2^{BS}\U^t_{BS}\V_{RF}^{BS}\T_1^{BS}\V_{BB}^{BS}\s + \n,
\end{equation}
An example of the summation array $\U^t_{BS}$ for $N_{ant}^{BS} =4 $ and $N_{RF}^{BS}=2 $ (i.e. 8 phase shifters) has the following structure:
\begin{equation}
\U^t_{BS} = 
\begin{bmatrix}
1 & 0 & 0 & 0 & 1 & 0 & 0 & 0\\ 
0 & 1 & 0 & 0 & 0 & 1 & 0 & 0\\
0 & 0 & 1 & 0 & 0 & 0 & 1 & 0\\
0 & 0 & 0 & 1 & 0 & 0 & 0 & 1\\
\end{bmatrix}
\label{eqn:sum_matrix_example}
\end{equation}

As $\U^t_{BS}$ can be viewed as a block row vector composed of $N_{RF}^{BS}$ identity matrix $\I_{N_{ant}^{BS}}$, i.e. $\begin{bmatrix}\U^t_{BS} = \I_{N_{ant}^{BS}} & \I_{N_{ant}^{BS}} & \cdots & \I_{N_{ant}^{BS}} \end{bmatrix}$, we can use a Kronecker product to commute $\T_2^{BS}\U^t_{BS}$ such as $\T_2^{BS}\U^t_{BS} = \U^t_{BS}(\I_{N_{RF}^{BS}}\otimes\T_2^{BS})$. This is equivalent to move the replicas of the PAs (as well as other components) near the transmit antennas onto each branch before the summation operation. A similar approach can be adopted for the UE, we can thus get an equivalent system model of \eqref{eqn:sig_model_full}:
\begin{equation}
\label{eqn:sig_model_full_equ}
\begin{aligned}
\y = &\underbrace{\W_{BB}^{UE}\W_{RF}^{UE}}_{\W_{UE}}
\underbrace{(\R_1^{UE}\otimes\I_{N_{ant}^{UE}})(\I_{N_{RF}^{UE}}\otimes\R_2^{UE})}_{\R_{UE}}
 \underbrace{\U^r_{UE}\C\U^t_{BS}}_{\tilde{\C}}\\
 &\underbrace{(\I_{N_{RF}^{BS}}\otimes\T_2^{BS})(\T_1^{BS}\otimes\I_{N_{ant}^{BS}})}_{\T_{BS}}
\underbrace{\V_{RF}^{BS}\V_{BB}^{BS}}_{\V_{BS}}\s + \n,
\end{aligned}
\end{equation}
where $\I_{N_{ant}^{BS}}$ and $\I_{N_{RF}^{UE}}$ are identity matrices. If we consider $\U^r_{UE}\C\U^t_{BS}$ as a composite propagation channel $\tilde{\C}$, the equivalent signal model is similar to \eqref{eqn:sig_equ_model}. 

When the system is in the UL transmission, the switches at the BS are connected to receive paths whereas those at the UE are connected to transmit paths. Thus, the UL composite channel can be written as $\U^r_{BS}\C^T\U^t_{UE}$, which can be verified as $\tilde{\C}^T$, implying that reciprocity is maintained for the composite propagation channel. Note that if there exists some hardware impairment in the summation operation, we can represent $\U^t$ and $\U^r$ by $\E^t\U_0^t$ or $\U_0^r\E^r$ where $\U_0$ is the ideal summation matrix as in \eqref{eqn:sum_matrix_example}, $\E^t$ and $\E^r$ are impairment matrices which can be absorbed into $\T_2^{BS}$ or $\R_2^{UE}$.

%The internal reciprocity calibration on for a fully connected architecture is similar to the subarray architecture. The full antenna array can be divided into two antenna groups $\cA$ and $\cB$. When the switches for $\cA$ are connected to the transmit path, antennas in $\cB$ are connected to receive path, and vise versa. Replacing $\C$ in \eqref{eqn:bi_trans} by the composite channel $\C'$ and all derivations are the same.

For a fully connected architecture, internal reciprocity calibration is not feasible since it is not possible to partition the whole antenna array into transmit and receive antenna groups. To enable TDD reciprocity calibration, a reference UE with a good enough channel should be selected to assist the BS to calibrate, such as \cite{shi2011efficient} proposed for a fully digital system. In this case, the bi-directional transmission no longer happens between two partitioned antenna groups $\cA$ and $\cB$ but between the BS and the UE. The selected reference UE needs to feed back its measured DL channel to the BS during the calibration procedure. Methods in Section~\ref{subsec:intern_calib} can still be used to estimate the calibration matrices for both BS and UE. Note that although UE feedback is heavy, the calibration does not have to be done very frequently, thus such a method is still feasible. Another possible way is to use a dedicated device at the BS to assist the antenna array for calibration, e.g., using a reference antenna as in \cite{shepard2012argos}. Using this method, DL channel measurements feedback from UE can be avoided, but a dedicated digital chain needs to be allocated to the assistant device, introducing an extra cost.

%--------------------------------- ---------------
\section{Simulation Results} 
%------------------------------------------------
As a proof-of-concept, we perform internal calibration simulation for a subarray hybrid transceiver with 64 antennas and 8 RF chains. To the extent of our knowledge, signal mixers and amplifiers are the main sources of hardware asymmetry. For different RF chains, signal mixers introduce random phases when multiplying the baseband signal with the carrier, whereas the gain imbalance between different amplifiers can cause their output signal having different amplitudes. Other components can also have some minor impacts, e.g., the non-accuracy in the phase shifter can add a further random factor to the phase. In this simulation, we capture the main effects of these hardware properties introduced by signal mixers and amplifiers, though the calibration method is not limited to this simplified case. We assume that the random phases introduced by the signal mixers in $\T_1^{BS}$ and $\R_1^{BS}$ are uniformly distributed between $-\pi$ and $\pi$ whereas the amplitudes in $\T_2^{BS}$ and $\R_2^{BS}$ are independent variables uniformly distributed in $[1-\epsilon \;1+\epsilon]$, with $\epsilon$ chosen such that the standard deviation of the squared-magnitude is 0.1.

The intra-array channel model between antenna elements strongly depends on the antenna arrangement in the array, antenna installation, as well as the frequency band. In the simulation, we focus on a sub-6GHz scenario and adopt the experiment based intra-array radio channel in \cite{vieira2017reciprocity}, where the physical channel $c_{i,j}$ between two antenna elements $i$ and $j$ in the same planar antenna array is modeled as
\begin{equation}
c_{i,j} = |\bar{c}_{i,j}|\mbox{exp}(j2\pi\phi_{i,j})+\tilde{c}_{i,j}.
\label{eqn:intra_array_ch_model}
\end{equation}
In \eqref{eqn:intra_array_ch_model}, $\bar{c}_{i,j}$ is the near field path\footnote{This term is called ``antenna mutual coupling" in \cite{vieira2017reciprocity}, which is slightly different from the classical mutual coupling defined in \cite{balanis2016antenna} where two nearby antennas are both transmitting or receiving. We thus call this term ``near field path" describing the main signal propagation from one antenna to its neighbor element.} between two antenna elements and $\tilde{c}_{i,j}$ absorbs all other multi-path contributions due to reflections from obstacles around the antenna array. For simplicity reasons, we assume the 64 antennas follows a co-polarized linear arrangement with an antenna space of half of the wavelength. According to the measurements in \cite{vieira2017reciprocity}, the magnitude for two half-wavelength spaced antennas are $-15$dB and at each distance increase of half of the wavelength, $|\bar{c}_{i,j}|$ decreases by 3.5dB. $\phi_{i,j}$ is modeled as uniformly distributed in $[0, 1)$ since a clear dependence with distance was not found. The multi-path component is modeled by an i.i.d zero-mean circularly symmetric complex Gaussian random variable with variance $\sigma^2=0.001$.

\begin{figure}[!t]
\centering
\begin{subfigure}{\columnwidth}
  \centering
  \includegraphics[width=\columnwidth]{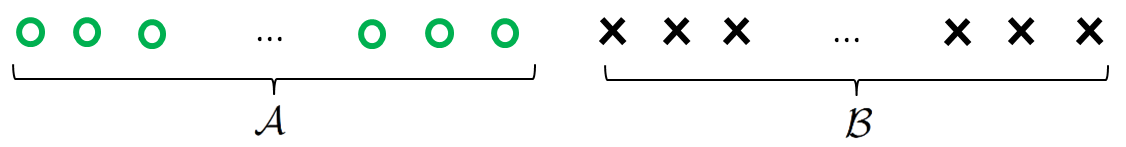}
  \caption{Two sides partition}
  \label{fig:two-sides-partition}
  \includegraphics[width=\columnwidth]{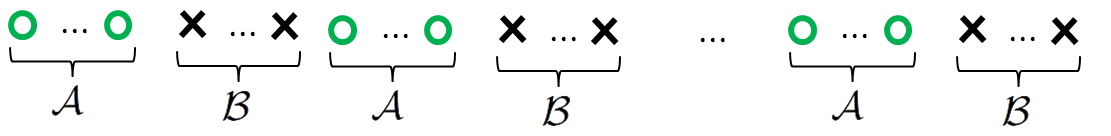}
  \caption{Interleaved Partition}
  \label{fig:interleaved-partition}
\end{subfigure}
\caption{(a) ``two sides partition" where group $\cA$ and $\cB$ contain 32 antennas on the left and right sides of the linear antenna array, respectively; (b) ``interleaved partition" where every 8 antennas are assigned to group $\cA$ and $\cB$, alternatively.}
\label{fig:antenna_partition}
\vspace{-1.0em}
\end{figure}

For the internal calibration, different antenna partition strategies are possible, where the optimal solution is yet to be discovered. In our simulation, we choose two different antenna partition scenarios: ``two sides partition" and ``interleaved partition", as shown in Fig.~\ref{fig:antenna_partition}. The ``two sides partition" separates the whole antenna array to group $\cA$ and $\cB$ on the left and right sides whereas the ``interleaved partition" assigns every 8 antennas to $\cA$ and $\cB$ alternatively.

%In this simulation, we use a Rician channel with $\kappa=0.7$, where the Rayleigh component in channel can be considered to be created by reflections from obstacles around the antenna array. For the simplicity of simulation, we also assume the same signal-to-noise ratio (SNR) of 40dB for all internal channels.
%This can be justified by the fact that as the antenna elements are quite close to each other, the main noise can come from the transmit antennas.
%In the calibration, we equally separate the whole antenna array into group $\cA$ and $\cB$. The entities in $\V_{BB}$ and $\W_{BB}$ are i.i.d Gaussian random variables. The phases in $\V_{RF}$ and $\W_{RF}$ are generated by a random uniform distribution in $[-\pi\;\pi]$. We accumulate 1600 measurements by setting $K=L=40$. 
%For the purpose of illustration, we do not assume $f_1=1$ as in Section \ref{subsec:intern_calib}, instead, we use the real value of $f_1$ as the reference. 

\begin{figure}[t]
\centering
\includegraphics[width=\columnwidth]{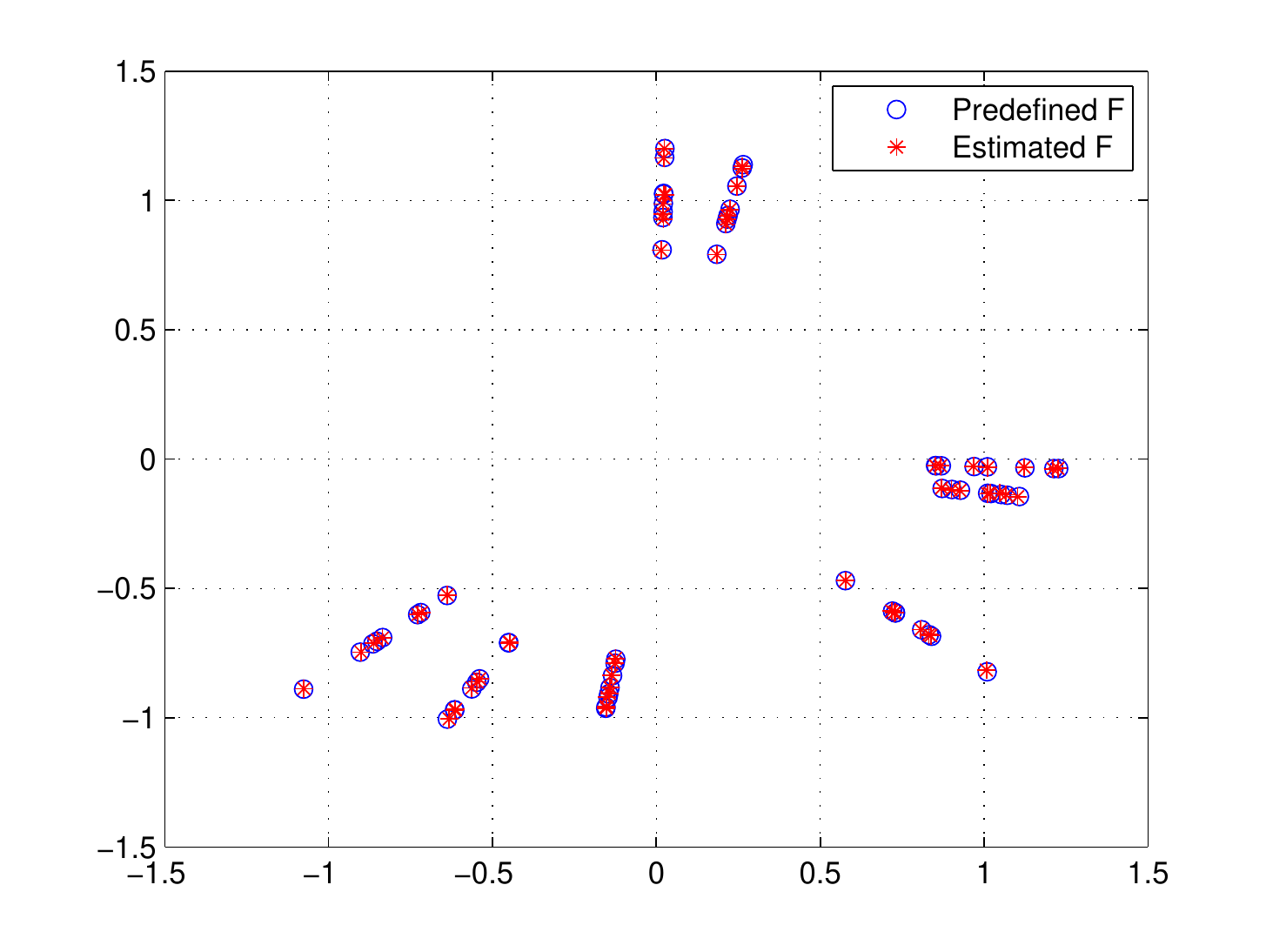}
\caption{Estimated calibration matrix vs. real calibration matrix. The blue circles are predefined calibration coefficients and the red stars are estimated values after elimination of the complex scalar ambiguity.}
\label{fig:F_est}
\vspace{-1.0em}
\end{figure}

In the first simulation, we would like to verify the feasibility to calibrate a hybrid beamforming transceiver using internal calibration. For this purpose, we use the ``two sides partition" scenario and assume no noise in the bi-directional transmission between group $\cA$ and $\cB$. We use 8 randomly generated independent QPSK symbols as pilots after the baseband digital beamforming and only apply analog precoding whose weights have a unit amplitude, with their phases uniformly distributed in $[-\pi\;\pi)$. Using $K=32$ and $L=5$ such randomly generated transmit and receive analog beam weights to accumulate 160 measurements\footnote{Note that in a practical multi-carrier system, the channel estimation on different RF chains can be performed on different frequencies as explained in Section~\ref{subsec:eff_ch_est}, the needed K can then be much less.} and applying the method \ref{subsec:intern_calib} on the accumulated signal, we can obtain the estimated calibration coefficients. For the purpose of illustration, we eliminate the complex scalar ambiguity, the results are shown in Fig.~\ref{fig:F_est}.
%Problems

We observe that the calibration matrix are partitioned in 8 groups, corresponding to 8 RF chains each with its own signal mixer. On each angle, elements have different amplitudes, which mainly correspond to the gain imbalance of independent amplifiers on each branch. We also observe that the estimated calibration parameters perfectly match the predefined values, implying that we can recover the coefficients using the proposed method. In a practical system, as no real value of $\F$ is known, all estimated coefficients have an ambiguity up to a common complex scalar value as explained in Section~\ref{subsec:intern_calib}.
%When $f_1$ is unknown in practice and assumed to be $1$, all estimated coefficients have an ambiguity up to a common complex scalar value as explained in Section~\ref{subsec:intern_calib}.

In the next simulation, we study the calibration performance with regard to the number of intra-array channel measurements. Since the measurements are within the antenna array, noise from both transmit and receive hardware can impact the received signal's quality. For antennas near each other, the main noise source comes from the transmit signal, usually measured in error vector magnitude (EVM). Assuming a transmitter with an EVM of $-20$dB, the SNR of the transmit signal is $40$dB. 
%\cite{shafik2006extended}.
For antennas far away from each other, noise at the receiver is the main limiting factor. Assuming that the system bandwidth is $5$MHz, the thermal noise at room temperature would be $-107$dBm at the receiving antenna. Using a radio chain with a noise figure of $10$dB and a total receive gain equaling to $0$dB, the noise received in the digital domain would be around $-97$dBm. We assume a 0dBm transmission power per antenna and use the intra-array channel model as in \eqref{eqn:intra_array_ch_model}. The calibrated coefficients are measured in its normalized mean square error (NMSE), such as 
\begin{equation}
\mbox{NMSE}_{\F} = \frac{\|\hat{\F}-\F\|^2}{\|\F\|^2}.
\end{equation}

%\begin{figure}[it]
%\centering
%\includegraphics[width=\columnwidth]{figures/mse_KL_iter1000_twosides_all}
%\caption{MSE of estimated calibration matrix vs. the number of K and L in the ``two sides partition scenario". Both Tx and Rx noise are considered.}
%\label{fig:MSE_F_two_sides}
%\end{figure}
%
%\begin{figure}[it]
%\centering
%\includegraphics[width=\columnwidth]{figures/mse_KL_iter1000_interleaved_all}
%\caption{MSE of estimated calibration vs. the number of K and L in the ``interleaved partition scenario". Both Tx and Rx noise are considered.}
%\label{fig:MSE_F_interleaved}
%\end{figure}
%
%\begin{figure}[it]
%\centering
%\includegraphics[width=\columnwidth]{figures/mse_KL_iter1000_twosides_trxN}
%\caption{MSE of estimated calibration vs. the number of K and L in the ``two sides partition scenario". Tx and Rx noise are simulated independently.}
%\label{fig:mse_KL_iter100_twosides_trxN}
%\end{figure}
%
%\begin{figure}[it]
%\centering
%\includegraphics[width=\columnwidth]{figures/mse_KL_iter1000_interleaved_trxN}
%\caption{MSE of estimated calibration matrix vs. the number of K and L in the ``interleaved partition scenario". Tx and Rx noise are simulated independently.}
%\label{fig:mse_KL_iter100_interleaved_trxN}
%\end{figure}

\begin{figure*}[!t]
\centering
\begin{subfigure}{\columnwidth}
\centering
\includegraphics[width=0.97\columnwidth]{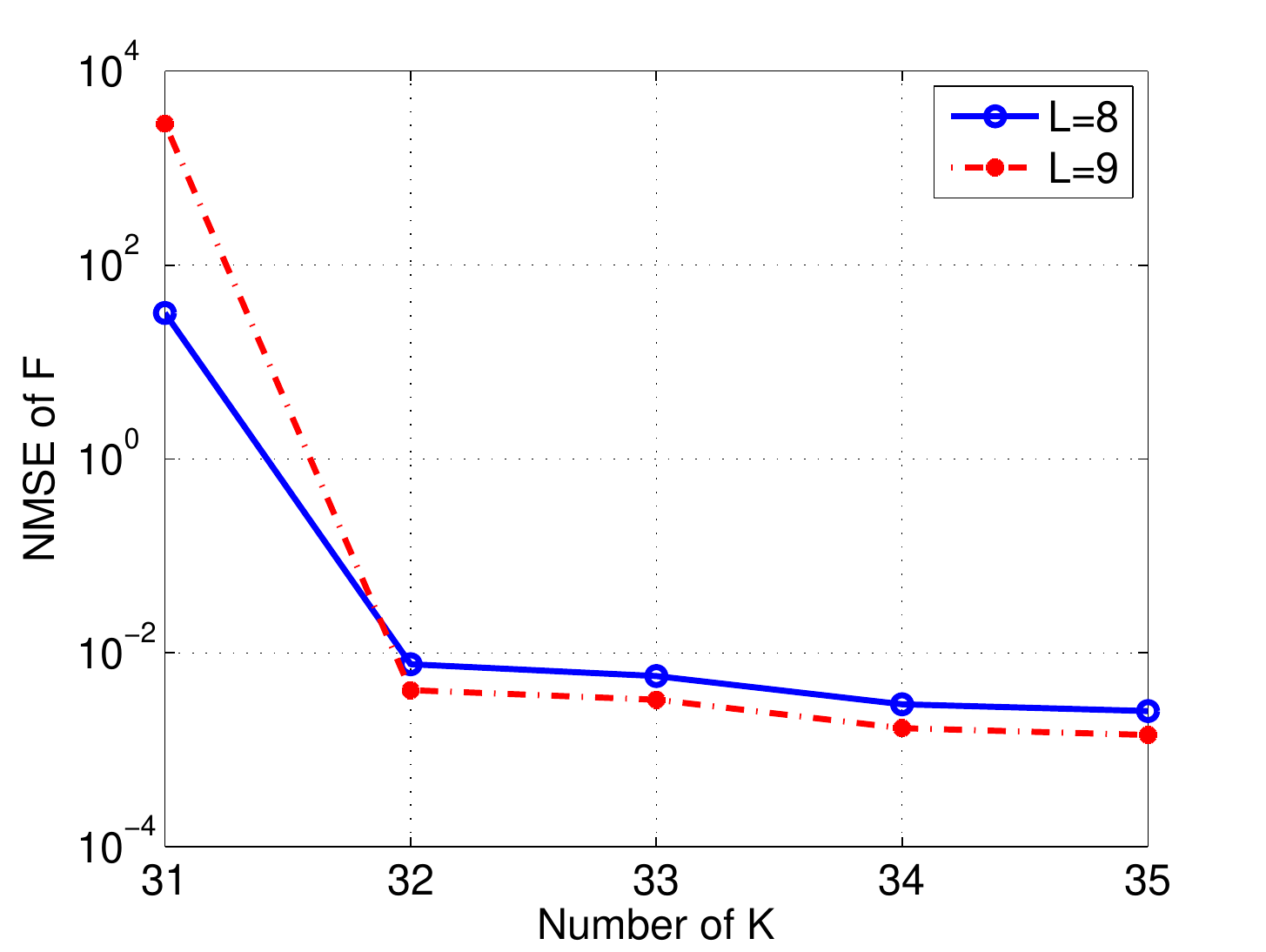}
\end{subfigure}
\begin{subfigure}{\columnwidth}
\centering
\includegraphics[width=0.97\columnwidth]{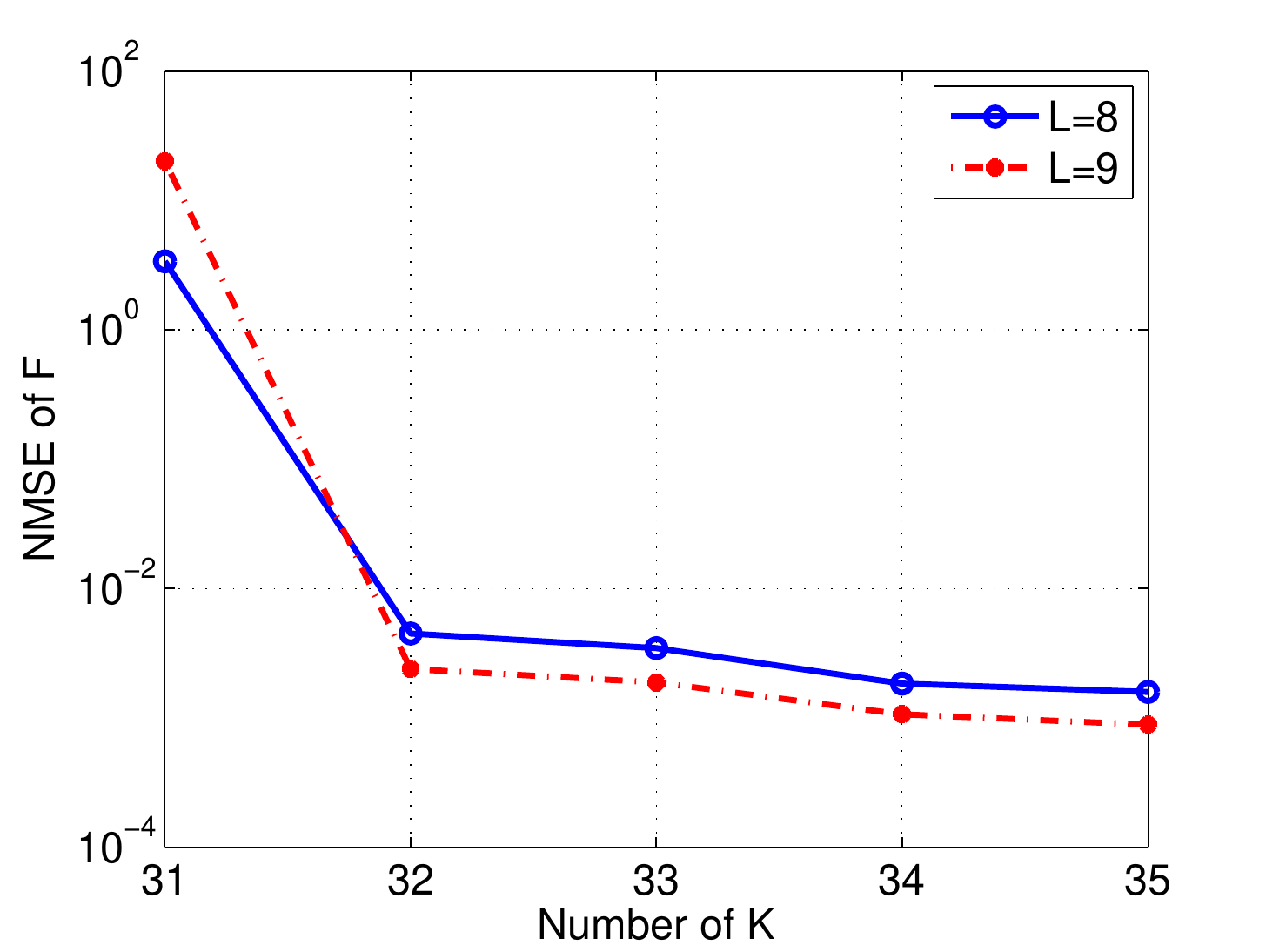}
\end{subfigure}
\caption{MSE of estimated calibration matrix vs. the number of K and L in (a) the ``two sides partition scenario" and (b) the ``interleaved partition scenario". Both Tx and Rx noise are considered.}
\label{fig:MSE_F}
\centering
\begin{subfigure}{\columnwidth}
\centering
\includegraphics[width=0.97\columnwidth]{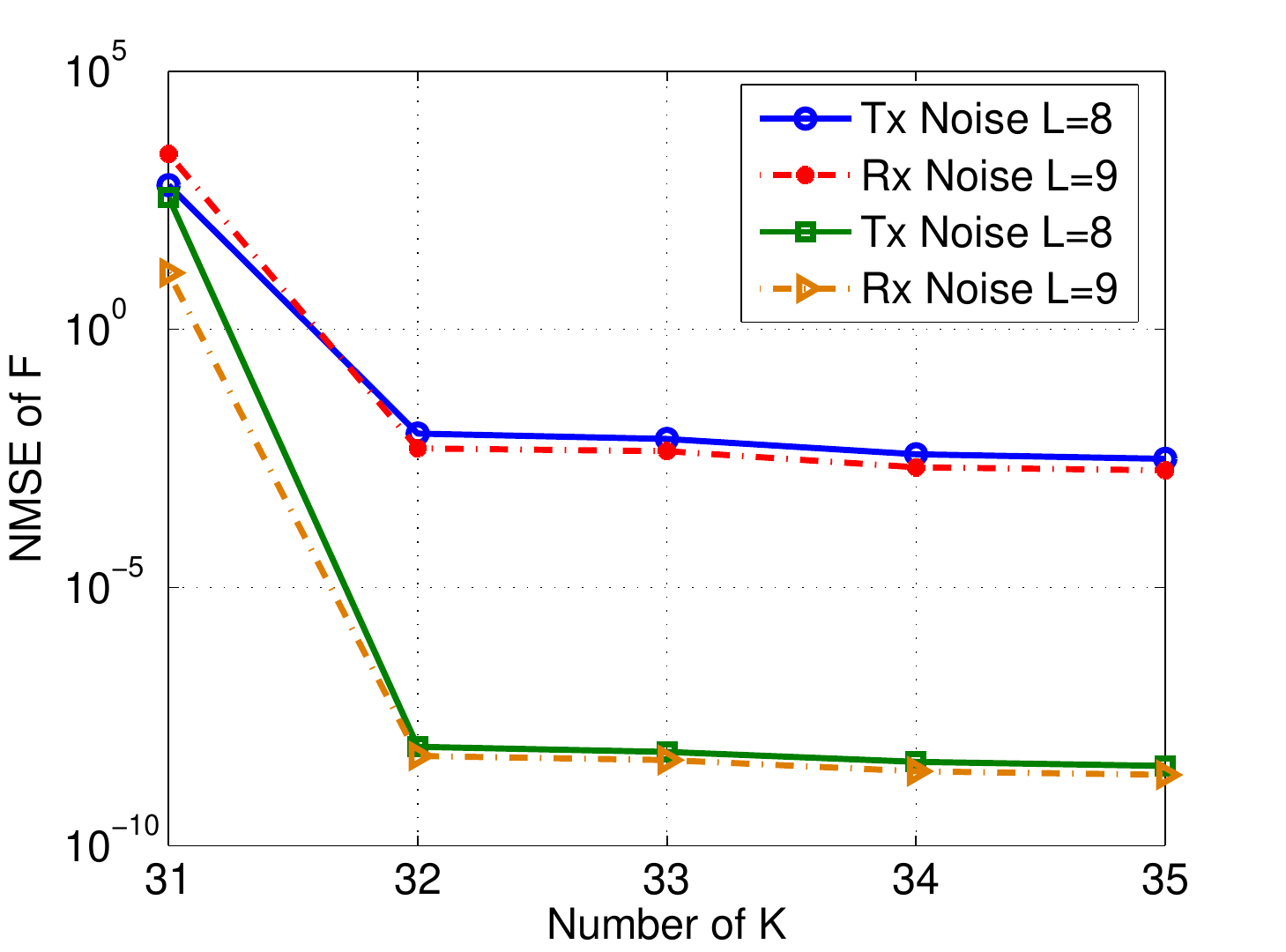}
\end{subfigure}
\begin{subfigure}{\columnwidth}
\centering
\includegraphics[width=0.97\columnwidth]{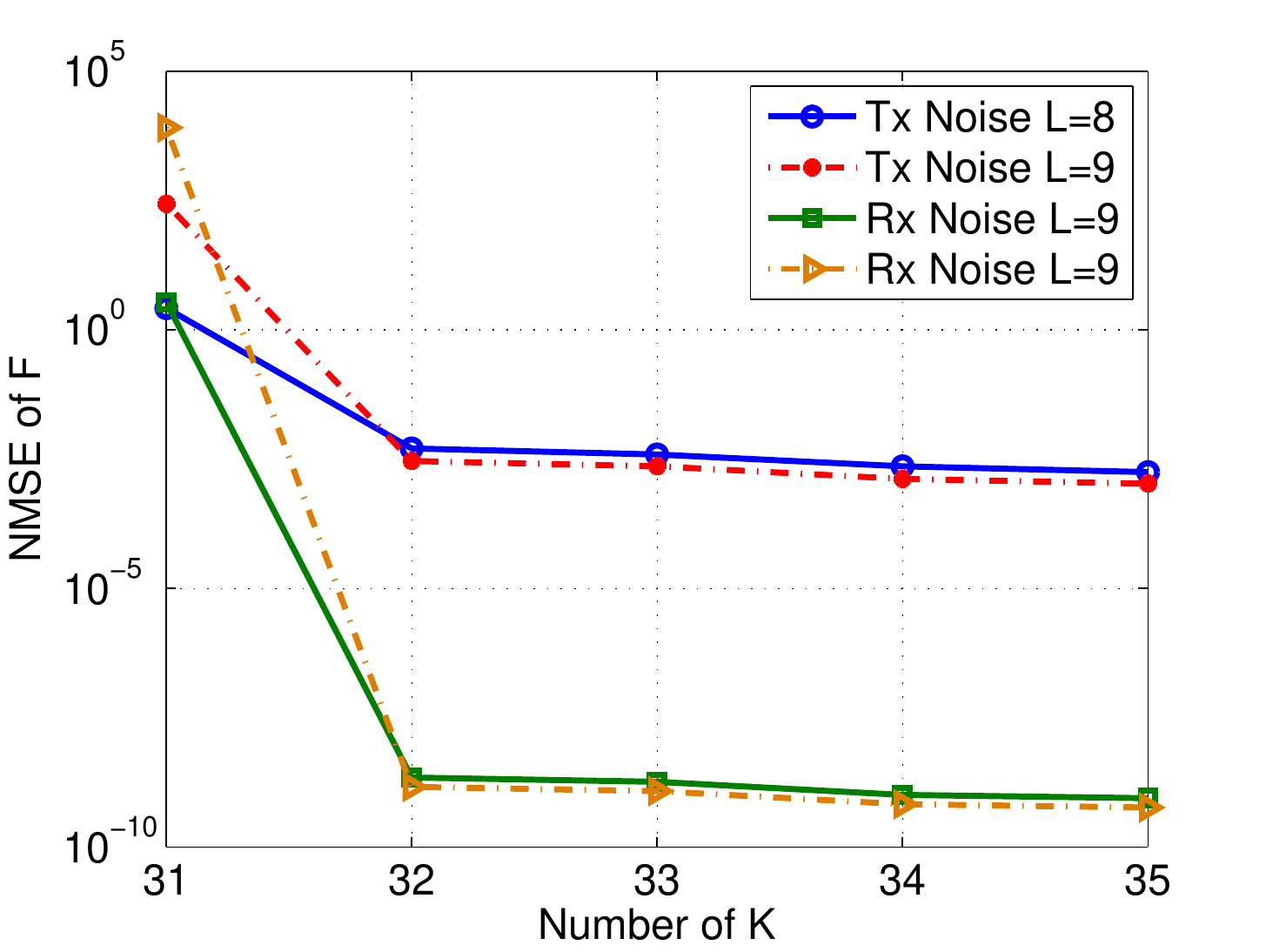}
\end{subfigure}
\caption{MSE of estimated calibration matrix vs. the number of K and L in (a) the ``two sides partition scenario" and (b) the ``interleaved partition scenario" . Tx and Rx noise are simulated independently.}
\label{fig:mse_trxN}
\vspace{-1.0em}
\end{figure*}

The results are shown in Fig.~\ref{fig:MSE_F} for ``two sides partition" and ``interleaved partition". We observe in both cases that, when $K < 32$, the estimation of $\F$ can not converge, since the intra-array channel estimation problem is under-determined, as explained in Section~\ref{subsec:eff_ch_est}. As long as $K \geq 32$ and $L \geq 8$, it is possible to estimate $\F$ up to an accuracy with an NMSE below $10^{-2}$. The ``interleaved partition" has a better performance than the ``two sides partition" when the minimum $K$ and $L$ requirements are met. This can be explained by the fact that the received signals in the ``interleaved partition" have more balanced amplitudes than in the ``two sides partition", where, the bi-directional transmission between far away antenna elements have very little impact on the estimation of $\F$ since the received signal are small.
Note that different sets of transmit and receive analog precoding weights can lead to different performance in the estimation of $\F$, with the best set left to be discovered. In our simulation, we randomly choose a set of weights and use it for both the ``two sides partition" and the ``interleaved partition". For comparison purpose, the set of weights for given $K$ and $L$ values (e.g $K=32, L=8$) is a subset for the weights used when $K$ and $L$ are bigger (e.g $K=33, L=9$).

Since we simulate the intra-array transmission, both the transmit and receive noise have been taken into account. In order to understand the impact from the two noise sources, let us simulate for them independently under both antenna partition scenarios. Fig.~\ref{fig:mse_trxN} illustrates the NMSE of $\F$ with independently considered noise for ``two sides partition" and ``interleaved partition". It is obvious that, in both cases, the noise at the transmit side is dominant and limits the accuracy of the estimated $\F$ whereas if only the receiver's thermal noise is considered, NMSE of $\F$ becomes negligible. In fact, if we look back at \eqref{eqn:LS_cost_fun}, it is the errors present in the bi-directional channel estimation $h_i$ and $h_j$ with the highest amplitudes (i.e. internal channels between nearby antenna elements) that dominate the cost function. For a receiving antenna near the transmitting element, the received transmit noise is much higher than the thermal noise generated at the receiving antenna itself.

When the system has accomplished internal calibration, it can use the estimated calibration matrix together with instantaneously estimated UL channel to assess the DL CSIT in order to create a beam for data transmission. The accuracy of the estimated DL CSIT depends on both the UL CSI and the estimated calibration matrices. In order to study the impact of both factors, we assume a simple scenario where a subarray hybrid structure BS performs beamforming towards a single antenna UE, such as in \cite{jiang2016accurately}. In this case, the DL channel $\h_{BS \rw UE}^T$ (we use transpose since the DL channel is a row vector) can be estimated by $\hat{f}_{UE}^{-1}\hat{\h}_{UE\rw BS}^T\hat{\F}_{BS}$, where $\hat{\h}_{UE\rw BS}$ is the estimated UL channel.  $\hat{\h}_{UE\rw BS} = \h_{UE\rw BS} + \Delta  \h_{UE\rw BS}$, where $\Delta  \h_{UE\rw BS}$ is the UL channel estimation error,  $\h_{UE\rw BS} = \R_{BS}\mathbf{c} t_{UE}$, with the UL physical channel vector $\mathbf{c}$ modeled as a standard Rayleigh fading channel. 
%$\hat{f}_{UE}^{-1}$ and $\hat{\F}_{BS}$ are the calibration coefficients for both UE and BS. Let us use $\F = f_{UE}\F_{BS}$ to denote the calibration matrix including coefficients from both UE and BS.
In our case, the calibration coefficients at the BS and the UE can be combined such as $\F = f_{UE}^{-1}\F_{BS}$.
Its estimation $\hat{\F}$ can be represented by $\hat{\F} = \F+\Delta \F$ with $\Delta \F$ denoting the estimation error. The estimation errors in $\Delta \h_{UE}$ and $\Delta \F$ are assumed to be i.i.d Gaussian random variables with zero mean and $\sigma_{n,UL}^2$, $\sigma_{\F}^2$ as their variance, respectively. $\mbox{NMSE}_{\F}$ can be calculated as $N_{ant}^{BS}\sigma_{\F}^2/\|\F\|^2$. Without considering the complex scalar ambiguity, which does not harm the finally created beam, we can calculate the NMSE of the DL CSI as
\begin{equation}
\label{eqn:mse_F_err}
\begin{aligned}
\mbox{NMSE}_{DL} =& 
\frac{1}{N_{ant}^{BS}}\mathbb{E}{\left[\|\hat{\h}_{UE\rw BS}^T\hat{\F}-\h_{BS\rw UE}^T\|^2\right]} \\
=& \frac{1}{N_{ant}^{BS}}\mathbb{E}{\left[\|\h_{UE \rw BS}^T\Delta \F + \Delta \h_{UE \rw BS}^T\hat{\F}\|^2\right]} \\
=&\frac{1}{N_{ant}^{BS}}\mbox{Tr}\left\{\Delta\F^H\mathbf{\Omega}^*\Delta\F + \sigma_{n, UL}^2\hat{\F}^H\hat{\F}\right\}
\end{aligned}
\end{equation}
where $\mathbf{\Omega}$ is the covariance matrix of the UL channel, i.e. $\mathbf{\Omega} = \mathbb{E}[\h_{UE \rw BS}\h_{UE \rw BS}^H]$.

\begin{figure}[!t]
\centering
\includegraphics[width=\columnwidth]{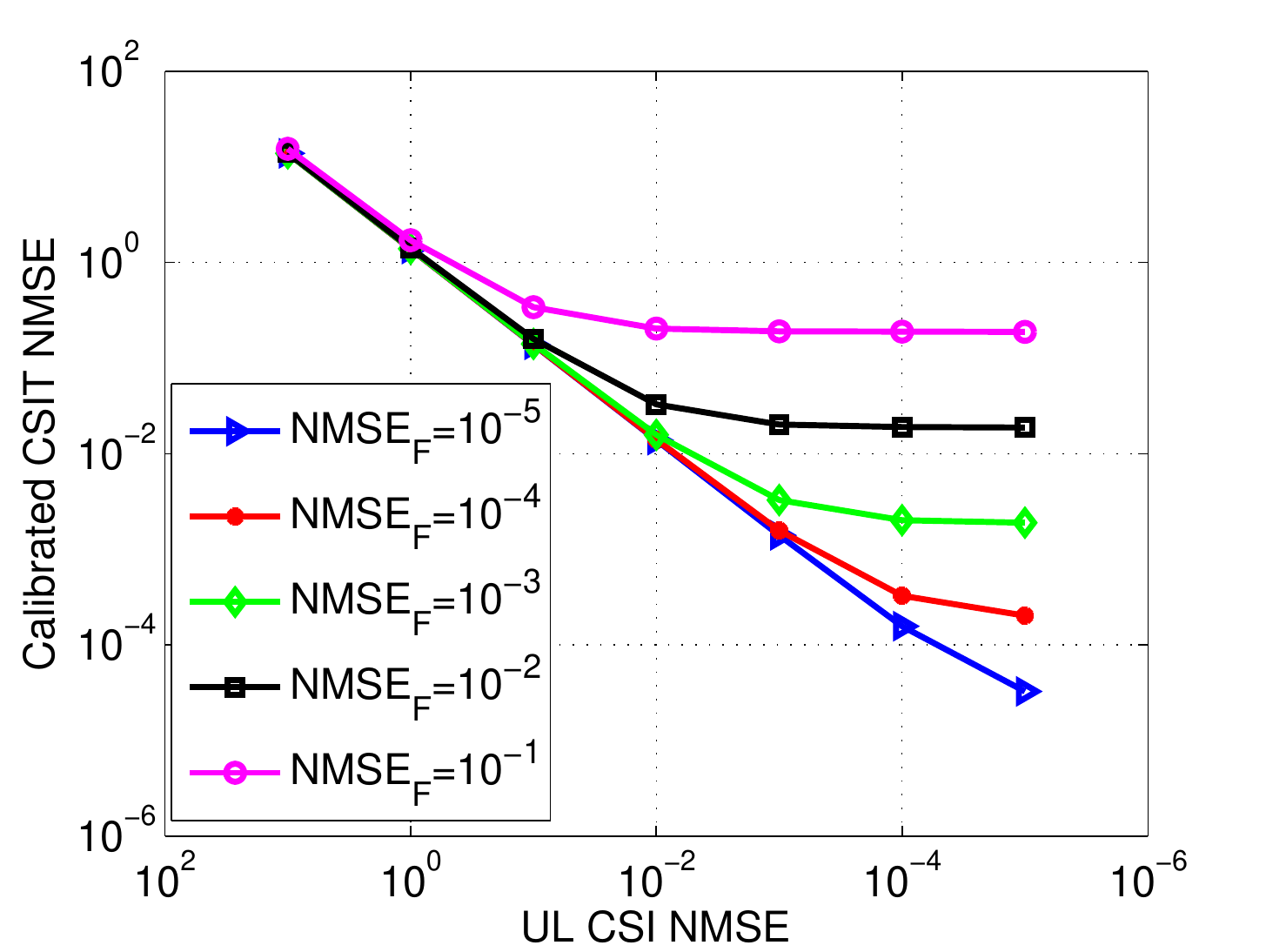}
\caption{The accuracy of acquired CSIT as a function of NMSE of the reciprocity calibration matrix and instantaneously measured UL CSI.}
\label{fig:MSE_vs_ULMSE}
\vspace{-1.0em}
\end{figure}

The NMSE of the calibrated CSIT as a function of different $\mbox{NMSE}_\F$ and $\mbox{NMSE}_{UL}$%\footnote{$\mbox{NMSE}_{UL}=\mathbb{E}\left[\frac{\|\Delta \h_{UE \rw BS}\|^2}{\|\h_{UE \rw BS}\|^2}\right] = \sigma_{n,UL}^2$.} 
\footnote{$\mbox{NMSE}_{UL}=\frac{1}{N_{ant}^{BS}}\mathbb{E}\left[\|\Delta \h_{UE \rw BS}\|^2\right] = \sigma_{n,UL}^2$.} 
is shown in Fig.~\ref{fig:MSE_vs_ULMSE}. We observe that when the accuracy of the UL CSI is low, it is the main accuracy limiting factor on the calibrated DL CSIT. As the UL CSI accuracy increases, the accuracy on $\hat{\F}$ begins to influence the DL CSIT. In a calibrated system where $\mbox{NMSE}_\F = 10^{-2}$ and $\mbox{NMSE}_{UL}=10^{-2}$, it is possible to obtain DL CSIT with an NMSE under $10^{-1}$.

%As $\F$ is quite stable during a relatively long time \cite{shepard2012argos} and the intra-array channels have quite high SNR, it is possible to use large enough $K$ and $L$ during the system initialization phase to make the estimation error on $\F$ very small. Thus, as long as we also have accurate enough UL channel estimation, we can use \eqref{eqn:H_recip} to obtain perfect DL CSIT, based on which the actual digital and analogue beamforming weights can be computed using, for example, the method presented in \cite{sohrabi2016hybrid2}.

%------------------------------------------------
\section{Conclusion} 
%------------------------------------------------
In this paper, we presented a CSIT acquisition method based on reciprocity calibration in a TDD hybrid beamforming massive MIMO system. Compared to state-of-the-art methods which assume a certain structure in the channel such as the limited scattering property validated only in mmWave, this method can be used for all frequency bands and arbitrary channels.
%This a clear advantage over state-of-the art methods based on certain assumptions on the structure of the channel, which are difficult to fulfill in practice and introduce a systematic loss in performance. 
Once the TDD system is calibrated, accurate CSIT can be directly obtained from the reverse channel estimation, without any beam training or selection. It thus offers a new way to operate hybrid AD beamforming systems.

%\begin{figure}[it]
%\centering
%\includegraphics[width=\columnwidth]{figures/mse}
%\caption{Up-most CSIT accuracy comparison of reciprocity and ray based methods}
%\label{fig:ch_est}
%\end{figure}

%=============== Acknowledgments ===============
%\section*{Acknowledgments}
%We acknowledge the support of Huawei Mathematical and Algorithmic Sciences Lab in Paris through the project of ``Modeling, Calibrating and Exploiting Channel Reciprocity for Massive MIMO". This work was also partly funded by the French Government (National Research Agency, ANR) through the ``Investments for the Future'' Program (\#ANR-11-LABX-0031-01). 

%=============== Bibliography ===============
\bibliographystyle{IEEEtran} 
\bibliography{refs}
\end{document}